\documentclass[utf8]{frontiersSCNS} 

\usepackage{url,hyperref,lineno,microtype,subcaption}
\usepackage[onehalfspacing]{setspace}
\usepackage{multirow}
\usepackage{subfloat}
\usepackage{rotating}
\usepackage{longtable}
\usepackage{tablefootnote}

\usepackage[symbol]{footmisc}

\def\kms{\,km\,s$^{-1}$}

\def\hb{{\sc{H}}$\beta$\/}

\def\feii{$\rm{Fe\;II}$}
\def\caii{\rm{Ca {\sc{ii}}}}
\def\rfe{R$_{\textmyfont{FeII}}$}

\def\lbol{$\mathrm{L_{bol}}$}
\def\mbh{$M\mathrm{_{BH}}$}

\def\mdot{$\dot{\mathcal{M}}$}

\def\LLEdd{$L\mathrm{_{bol}}/L\mathrm{_{Edd}}$}

\def\RL{$R\mathrm{_{H\beta}-L_{5100}}$}

\def\zsun{Z$_{\odot}$}
\def\msun{M$_{\odot}$}
\def\un{$\log \rm{U} - \log \rm{n_{H}}$}
\def\rblr{$R\mathrm{_{BLR}}$}

\def\hb{H$\beta$}

\def\n{$\rm{n_{H}}$}
\def\u{$\rm{U}$}
\def\l5100{$\mathrm{L_{5100}}$}
\def\kbol{$k\mathrm{_{bol}}$}

\newenvironment{myfont}{\fontfamily{lmtt}\selectfont}{\par}
\DeclareTextFontCommand{\textmyfont}{\myfont}

\def\keyFont{\fontsize{8}{11}\helveticabold }
\def\firstAuthorLast{Panda}

\def\Authors{Swayamtrupta Panda\,$^{1,2,3,\dagger, \thanks{CNPq Fellow}}$} 

\def\Address{$^{1}$Center for Theoretical Physics, Polish Academy of Sciences, Al. Lotnik{\'o}w 32/46, 02-668 Warsaw, Poland \\
$^{2}$Laborat\'orio Nacional de Astrof\'isica - MCTIC, R. dos Estados Unidos, 154 - Na\c{c}\~oes, Itajub\'a - MG, 37504-364, Brazil\\
$^{3}$Nicolaus Copernicus Astronomical Center, Polish Academy of Sciences, ul. Bartycka 18, 00-716 Warsaw, Poland
}

\def\corrAuthor{Swayamtrupta Panda}

\def\corrEmail{panda@cft.edu.pl}

\begin{document}
\onecolumn
\firstpage{1}

\title[\RL{} relations and Eigenvector 1]{Parameterizing the AGN radius - luminosity relation from the Eigenvector 1 viewpoint} 

\author[\firstAuthorLast ]{\Authors} 
\address{} 
\correspondance{} 

\extraAuth{}

\maketitle

\begin{abstract}

\section{}
The study of the broad-line region (BLR) using reverberation mapping has allowed us to establish an empirical relation between the size of this line emitting region and the continuum luminosity that drives the line emission (i.e. the \RL{} relation). To realize its full potential, the intrinsic scatter in the \RL{} relation needs to be understood better. The mass accretion rate (or equivalently the Eddington ratio) plays a key role in addressing this problem. On the other hand, the Eigenvector 1 schema has helped to reveal an almost clear connection between the Eddington ratio and the strength of the optical \feii{} emission that originates from the BLR. This paper aims to reveal the connection between theoretical entities, like, the ionization parameter (\u{}) and cloud mean density (\n{}) of the BLR, with physical observables obtained directly from the spectra, such as optical \feii{} strength (\rfe{}) that has shown immense potential to trace the accretion rate. We utilize the photoionization code CLOUDY and perform a suite of models to reveal the physical conditions in the low-ionization, dust-free, line emitting BLR. The key here is the focus on the recovery of the equivalent widths (EWs) for the two low-ionization emission lines - \hb{} and the optical \feii{}, in addition to the ratio of their EWs, i.e., \rfe{}. We compare the spectral energy distributions of prototypical Population A and Population B sources, {\sc I Zw 1} and {\sc NGC 5548}, respectively, in this study. The results from the photoionization modelling are then combined with existing reverberation mapped sources with observed \rfe{} estimates taken from the literature, allowing us to assess our analytical formulation to tie together the aforementioned quantities. The recovery of the correct physical conditions in the BLR then suggests that - the BLR ``sees'' only a very small fraction ($\sim$1-10\%) of the original ionizing continuum.


\tiny
 \keyFont{ \section{Keywords:} galaxies: active, quasars: emission lines; accretion -- reverberation mapping, scaling relations; photoionization modelling} 
\end{abstract}

\section{Introduction}
\label{sec:intro}

The study of the broad-line regions in active galaxies has a long and inspiring history. The first signs of the detection of such emitting regions were noticed by \citet{seyfert1943} using a sample of nearby, low-luminosity active galaxies, which became popular as Seyfert galaxies. Then came the seminal work by \citet{schmidt63} wherein he discovered quasars to be of extragalactic origin. He studied the optical spectrum of a bright radio galaxy - 3C 273 and noted that the source had a redshift, z $\sim$ 0.158, using the strong, broad Balmer lines that were found to be shifted redwards to the reference lab-frame spectrum. Another, equally important discovery was the discovery of the variation in the intensities of these emission lines over some time, especially in the timescales of weeks to months. This implied very small emitting regions, of the order of a few 10$^{3}$ Schwarzchild radii \citep{greenstein_schmidt64}. This emitting region is now well known as the broad-line region (BLR). This crucial discovery opened up a new sub-field in the form of reverberation mapping and led to the estimation of the black hole masses of over hundreds of low- to high luminosity Seyferts and quasars \citep{blandford_mckee82,peterson88,peterson93,peterson2004} supplemented by single/multi-epoch spectroscopy \citep{kaspi2000, bentz13, dupu2014}. As we can already notice, the location of the BLR (R$_{\rm BLR}$) is closely related to the continuum properties, one that is linked to the underlying accretion disk. The primary observable quantity among these properties is the luminosity of the source which was realised already in \citet[][and references therein]{kaspi2005}. Later studies \citep[e.g.,][]{bentz13} improved on the H$\beta$-based \RL{} by the inclusion of more sources and removing the contribution of the host galaxy from the total luminosity. There has been a significant increase in the monitoring of archival sources and inclusion of newer ones which has begun to show a significant scatter from the empirical \RL{} relation \citep{grier17, martinez-aldama2019, du2019, panda_2019_frontiers}. This scatter informs us that there is a subset of sources that are observed at relatively high luminosities (log L$_{5100}$ $\gtrsim$ 43.0, in erg s$^{-1}$) for which the reverberation mapping yields shorter time-lags, thus shorter R$_{\rm BLR}$, than expected from the empirically derived estimates. Studies have pointed out the link to the accretion rate that could factor into explaining this scatter and provided corrections to the empirical relation in terms of observables that trace the accretion rate, e.g., strength of the optical \feii{} emission \citep{du2019}.



\begin{figure}[!htb]
    \centering
    \includegraphics[width=\textwidth]{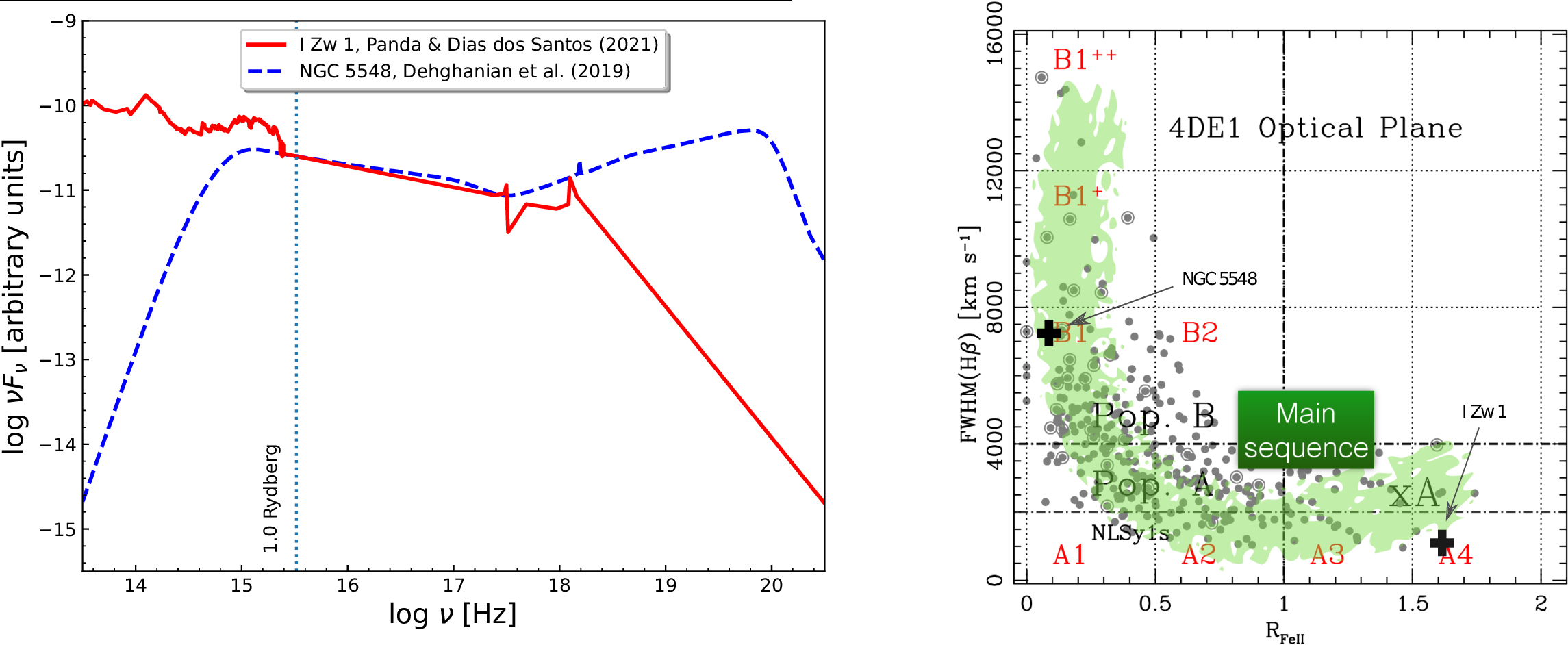}
    \caption{(left:) Comparison of the spectral energy distributions (SEDs) for \textmyfont{I Zw 1} (in red) and \textmyfont{NGC 5548} (in dashed blue). The \textmyfont{I Zw 1} SED is taken from \citet{panda_denimara_2021}, for \textmyfont{NGC 5548} we use the \citet{dehghanianetal2019} version of the \citet{mehdipouretal2015} SED. The SEDs are normalized at 1 Rydberg. (right:) Optical plane of the quasar main sequence (or 4DE1). Abridged version of \citet{mar18}. The location of \textmyfont{I Zw 1} (Marziani et al. 2021, submitted) and \textmyfont{NGC 5548} \citet{du2019} are marked with plus symbols.}
    \label{fig:sed}
\end{figure}

The spectral diversity of Type-1 AGNs was brought together under a single framework by the study of \citet{borosongreen1992}. The paper of \citet{borosongreen1992} is fundamental for two reasons: (A) It provided one of the first template for fitting the \feii{} pseudo-continuum\footnote{The \feii{} emission manifests as a pseudo-continuum owing to the many, blended multiplets over a wide wavelength range \citep[see][and references therein]{verner99,kovacevic2010}.} extracted from the spectrum of a prototypical Narrow Line Seyfert Type-1 (NLS1) source, {\sc I Zw 1}; and (B) for introducing the Main Sequence of Quasars to unify the diverse group of AGNs. They were among the first to use dimensionality reduction on observed properties of quasars to obtain this main sequence, specifically the Eigenvector 1 which eventually led to the connection between the FWHM of the broad H$\beta$ and the strength of the \feii{} blend between 4434-4684 \AA~ (i.e., the ratio of the EW(\feii{}) to the EW(H$\beta_{\rm broad}$), or more commonly known as \rfe{}). This is now the well established ``Quasar Main Sequence'' in the optical plane (see e.g., the right panel in Figure \ref{fig:sed}) which is found to be primarily driven by the Eddington ratio among other physical properties\footnote{Modelling the \feii{} pseudo-continuum requires the knowledge of 8-dimensional parameter space, one that encompasses the full diversity of Type-1 AGNs as has been concluded from prior works \citep{panda18b, panda19, panda19b, panda19c}. These 8 parameters consist of the fundamental black hole (BH) and BLR properties, namely (1) the Eddington ratio (\LLEdd{}); (2) the BH mass (\mbh{}), (3) the shape of the ionizing continuum or the spectral energy distribution (SED), (4) the BLR local density (\n{}), (5) the metal content in the BLR, (6) the velocity distribution of the BLR including turbulent motion within the BLR cloud, (7) the orientation of the source (as well as the BLR) to the distant observer, and (8) the sizes of the BLR clouds \citep[see][for a comprehensive review]{Panda_PhD}.} \citep[e.g.,][]{sul00,sh14,mar18,panda18b,panda19,panda19b}.

In addition to these developments, a classification based on the narrowness or broadness of the H$\beta$ emission line profile in an AGN spectrum was introduced, i.e., as Population A and Population B. Population A sources can be understood as the class that includes local NLS1s as well as more massive high accretors which are mostly classified as radio-quiet \citep[e.g.,][]{ms14} and that have FWHM(H$\beta$) $\lesssim$ 4000 km s$^{-1}$. Previous studies have found that the Population A sources typically have Lorentzian-like H$\beta$ profile shape \citep[e.g.,][]{sul02,zamfiretal10} in contrast to Population B sources, the latter are shown to have broader H$\beta$ ($\gtrsim$ 4000 km s$^{-1}$), are pre-dominantly ``jetted'' sources \citep[e.g.,][]{padovani2017} and have been shown to have H$\beta$ profiles that are a better fit with Gaussian (for sources with still higher FWHMs, we observe disk-like double Gaussian profiles in Balmer lines). The cut off in the FWHM of H$\beta$ at 4000 km s$^{-1}$ was suggested by \citet{sul00,mar18} who found that AGN properties appear to change more significantly at this broader line-width cutoff. Later studies revealed that the two populations rather form a smooth link and are related \citep{fb17,berton2020}. The shape of the emission line profiles and continuum strength and shape is directly connected to the central engine, especially to the black hole mass, and, the accretion rate, in addition to the black hole spin and the angle at which the central engine is viewed by a distant observer \citep{czerny2017, mar18, panda18b, panda19b, Panda_PhD}.

Another important factor in the context of line formation in the BLR is the ionizing continuum that is incident on the BLR and as a result, produces those emission lines that we see in an AGN spectrum. The study of the spectral energy distribution (SED) is a key element in understanding how the BLR responds to the continuum, and especially through the study of the emission lines, as a whole, be able to answer how much of this incoming radiation is intercepted by the BLR and how much of this intercepted radiation leads to the line-formation and emission \citep[see e.g.,][]{korista_goad_2004, ch11, marziani_atoms2019, 2019arXiv190800742C}. The characterization of the ionizing SED, the part of it that comes from regions closer than the BLR, is important for our study of the emission lines, especially that carry photon energy at or above 1 Rydberg. This threshold marks the minimum energy required to ionize neutral hydrogen. From the photoionization point of view, this fraction of the broad-band SED is closely related to the number of ionizing photons that eventually lead to the line production. \citet{wandel99,negrete2014,mar15} have used this method to estimate the photoionization radius of the line-emitting region of the BLR. 

\citet{negrete2014, mar15} have used line diagnostic ratios in the UV to infer the densities and ionization parameters, especially for the high-ionization line emitting regions in the BLR\footnote{ionization parameter (\u{}) is a dimensionless parameter that informs about the total number of ionizing photons available for photoionization of a medium at a given density (\n{}).} But we lack such direct diagnostics for the density and ionization parameter in the optical regime. The optical part of the AGN spectrum contains emission lines, e.g. H$\beta$, \feii{}, that belong to the class of the low-ionization lines, i.e. with ionization potential (IP$<$20 eV, \citealt{collin88, marziani_2019_atoms}) that are theorized to be produced at scales that are larger than the regions that emit the high-ionization lines, e.g. C {\sc iv}$\lambda$1549 or He {\sc ii}$\lambda$1640 \citep{jol87,martinez-aldamaetal15,murilo2016}. In \citet{panda2021} (see also \citealt{panda_denimara_2021}), we outlined a method to account for the line EWs of H$\beta$ and optical \feii{} in addition to their ratio (i.e. \rfe{}). This allows us to evaluate the appropriate physical conditions, primarily in terms of density, ionization parameter and metal content. This method also brings into agreement the radius estimated using photoionization method to that of the reverberation mapping for sources that are accreting at or below the Eddington limit, such that they agree with the empirical \RL{} relation. In this paper, we re-iterate on the formalism but incorporate the standard \RL{} \citep{bentz13} as well as the \rfe{}-dependent \RL{} relation \citep{du2019} to study the effect of the accretion rate-dependent \rfe{} on our existing inferences. We test our model by incorporating the spectral properties of a prototypical Population A source - \textmyfont{I Zw 1}, and a prototypical Population B source - \textmyfont{NGC 5548}, and assess how much fraction of the ionizing continuum actually leads to the low-ionization line formation and emission in the dust-free BLR. The location of the two sources on the main sequence diagram is shown in the right panel of Figure \ref{fig:sed}. 



The paper is organized as follows: We describe the analytical prescription in Section \ref{sec:analytical_description} to combine the information from the \RL{} relations into the photoionization theory accounting for different bolometric corrections. We outline our photoionization modelling setup in Section \ref{sec:cloudy}. We analyze the results obtained from our analyses highlighting the strengths and weaknesses of our current model in Section \ref{sec:results}, and discuss open issues in the context of our work in Section \ref{sec:discussions}. We summarize our findings from this study in Section \ref{sec:conclusions}. Throughout this work, we assume a standard cosmological model with $\Omega_{\Lambda}$ = 0.7, $\Omega_{m}$ = 0.3, and H$_0$ = 70 \kms{} Mpc$^{-1}$.

\section{Analytical description}
\label{sec:analytical_description}
In order to realise the parameter space for the BLR and to link the physical quantities (\u{}, \n{}) and the observables - the AGN continuum luminosity at 5100\AA\ (\l5100{} = 5100\AA*$L_{5100}$, where $L_{5100}$ is directly estimated from the observed spectrum) and later also with the strength of the optical \feii{} emission (i.e, \rfe{}), we present the analytical relationships as described in the following sub-sections. We separately show the relations based on (A) the \RL{} relation used, and (B) the format of the bolometric correction used to scale the \l5100 to the bolometric luminosity (\lbol{}). We use two instances of the \RL{} relation - (a) the classical \citet{bentz13} \RL{} relation wherein the separation between the continuum source and the onset of the BLR (\rblr{}) is dependent only on the continuum luminosity of the source; and (b) a new \RL{} relation that incorporates the dependence of the \rfe{} in addition to the \l5100{} on the \rblr{} \citep{du2019}. 

In order to scale the \l5100{} to obtain the corresponding bolometric luminosity (\lbol{}), we incorporate two formats of the bolometric correction (hereafter \kbol) factor - (i) a fixed value derived from the mean SED from \citet{richards2006}; and (ii) a variable factor that is dependent on the luminosity of the source \citep{netzer2019}. The value for the \citet{richards2006} \kbol = 9.26\footnote{this value is estimated at 5100\AA\ for the mean SED from \citet{richards2006}.} has been used widely in statistical studies for large quasar catalogues \citep{shen11,rakshitetal2020}. Here, \kbol\ scales with the monochromatic luminosity ($\mathrm{L_{5100}}$) to give a rough estimation of \lbol$=k_{\mathrm{bol}} \cdot \rm{L_{5100}}$. Usually, \kbol\ is taken as a constant for a monochromatic luminosity; however, results like the well-known non-linear relationship between the UV and X-ray luminosities \citep[e.g.][and references therein]{lusso-risaliti2016} indicate that \kbol\ should be a function of luminosity \citep{marconi2004, krawczyk2013}. Along the same line, \citet{netzer2019} proposed new bolometric correction factors as a function of the luminosity assuming an optically thick and geometrically thin accretion disk, over a large range of black hole mass ($10^{7}$- $10^{10}$ M$_\odot$), Eddington ratios (0.007-0.5), spin (-1 to 0.998) and a fixed disk inclination angle of $56^\circ$. For the optical range (at 5100\AA), the bolometric correction factor is given by:
\begin{align}
k_\mathrm{bol}=40\left(\frac{L_{\rm opt}}{10^{42}}\right)^{-0.2},
\label{equ:kbol}
\end{align}
which is taken from the Table 1 in \citealt{netzer2019}. Here, $L_{\rm opt}$ = \l5100{}. The wide option of parameters considered for the model process provide a better approximation corroborating previous results \citep{nemmen-brotherton2010, runnoe2012a, runnoe2012b}. In addition, it provides a better accuracy than the fixed bolometric factor correction which led to errors as large as $50\%$ for individual measurements. Therefore, we also explore the use of the two different \kbol\ - fixed and the luminosity-dependent versions, in our analyses.


We start with the conventional description of the ionization parameter,\\
\begin{equation}
    U = \frac{\Phi(H)}{n_{H} c} = \frac{Q(H)}{4\pi R_{BLR}^2 n_H c} 
    \label{eq1}
\end{equation}
where, \textit{$\Phi(H)$} is the surface flux of ionizing photons (in $\mathrm{cm^{-2}\;s^{-1}}$) and \textit{$\rm{n_H}$} is the total hydrogen density (in $\mathrm{cm^{-3}}$). \textit{Q(H)} is the number of hydrogen-ionizing photons emitted by the central object (in $\mathrm{s^{-1}}$) and \rblr{} is the separation between the central source of ionizing radiation and the inner face of the cloud (in cm).

The \textit{Q(H)} term in the above equation can then be replaced with the equivalent \textit{instantaneous} bolometric luminosity ($L_{bol}$),\\
\begin{equation}
 Q(H) = \chi\frac{L_{bol}}{h\nu} 
 \label{eq2}
\end{equation}
Here, we consider the average photon energy, $h\nu$ = 1 Rydberg\footnote{$\approx$ 2.18$\times10^{-11}$ erg.} \citep{wandel99, marz15}. Not all of the bolometric luminosity is used to ionize the BLR. Based on the average photon energy, we consider a fraction of the total luminosity, i.e. the ionizing luminosity (L$_{\rm ion}$). The coefficient $\chi$ accounts for this fraction. The exact value of this coefficient is dependent on the shape of the input SED.  In this work, we assume $\chi$ = 0.5, which is estimated for the default AGN SED in {\sc CLOUDY} \citep{mf87,f17}.

Combining Equations \ref{eq1} and \ref{eq2}, we have
\begin{equation}
    \log (Un_{H}) = \log (L_{bol}) - \log (2h\nu4\pi c) - 2\log(R_{BLR})
    \label{eq3}
\end{equation}

Next, we look into the classical \RL{}, i.e. from the \textit{Clean} sample of \cite{bentz13}, we have,\\
\begin{equation}
    \log \left ( \frac{R_{BLR}}{1\;light-day}\right )  = \kappa + \alpha \log\left ( \frac{\mathrm{L_{5100}}}{10^{44}}\right )
     \label{eq4}
\end{equation}
where $\mathrm{L_{5100}}$ is the monochromatic luminosity at 5100 \AA (in units of $10^{44}$ erg s$^{-1}$). $\kappa$ and $\alpha$ take the values 1.555$\pm$0.024 and 0.542$^{+0.027}_{-0.026}$ respectively for the \textit{Clean} sample \citep[see Table 14 in][]{bentz13}. Here, The \rblr{} is normalized to 1 light day\footnote{= 2.59$\times 10^{15}$ cm.}.

Substituting the \citet{richards2006} value for the \kbol{} (=9.26) and the form of \rblr{} (from Equation \ref{eq4}) in Equation \ref{eq3}, we have for the fixed \kbol{} case:
\begin{equation}
    \boxed{\log (Un_{H}) = 9.815 - 0.084\log\left(\frac{\rm{L_{5100}}}{10^{44}}\right)}
    \label{eq5}
\end{equation}


Substituting the \citet{netzer2019} relation (see Equation \ref{equ:kbol}) for the \kbol{} and the form of \rblr{} (from Equation \ref{eq4}) in Equation \ref{eq3}, we have for the luminosity-dependent \kbol{} case:
\begin{equation}
    \boxed{\log (Un_{H}) = 10.050 - 0.284\log\left(\frac{\rm{L_{5100}}}{10^{44}}\right)}
    \label{eq6}
\end{equation}


Now, alternatively, a new \RL{} relation has been proposed by \citet{du2019} wherein the authors have incorporated the dispersion noticed in the classical \RL{}, especially due to some sources that deviated from the standard relation because shorter time-lags from reverberation mapping were obtained for them \citep[see e.g., Figure 1 in][for a recent compilation of reverberation-mapped sources]{panda19c}. This aspect has been studied for quite some time and certain correction factors were suggested to alleviate the dispersion in order to keep the classical \RL{} relation intact., e.g. in \citet{martinez-aldama2019} we found that this dispersion can be accounted for in the standard \RL{} with an added dependence on the Eddington ratio (\LLEdd{}) or the dimensionless accretion rate parameter (\mdot{}). In their paper, \citet{du2019} propose a dependence on the \rfe{} parameter which can be viewed as a proxy of the accretion rate effect. Inclusion of the \rfe{} parameter in the \RL{} relation has substantially reduced the scatter in the existing relation, from 0.299 dex \citep[see][]{2021POBeo.100..287M} to $\sim$ 0.19 dex. We refer the readers to the figures 6 and 7 in their paper \citep{du2019} for a comparison between the classical \RL{} relation and the new \rfe{}-dependent \RL{} relation. The formalism of \citet{du2019} has the following form,

\begin{equation}
    \log \left ( \frac{R_{BLR}}{1\;light-day}\right )  = \kappa ' + \alpha '' \log\left ( \frac{\mathrm{L_{5100}}}{10^{44}}\right ) + \gamma ' R_\mathrm{\textmyfont{FeII}}
     \label{eq7}
\end{equation}

here, $\kappa '$ = 1.65 $\pm$ 0.06, $\alpha ''$ = 0.45 $\pm$ 0.03, and $\gamma '$ = -0.35 $\pm$ 0.08. Substituting the \citet{richards2006} value for the \kbol{} (=9.26) and the form of \rblr{} (from Equation \ref{eq7}) in Equation \ref{eq3}, we have for the fixed \kbol{} case:
\begin{equation}
    \boxed{\log (Un_{H}) = 9.625 + 0.1\log\left(\frac{\rm{L_{5100}}}{10^{44}}\right) + 0.7R_{\rm FeII}}
    \label{eq8}
\end{equation}


In the same manner as before, substituting the \citet{netzer2019} relation (see Equation \ref{equ:kbol}) for the \kbol{} and the form of \rblr{} (from Equation \ref{eq7}) in Equation \ref{eq3}, we have for the luminosity-dependent \kbol{} case:
\begin{equation}
    \boxed{\log (Un_{H}) = 9.860 - 0.1\log\left(\frac{\rm{L_{5100}}}{10^{44}}\right) + 0.7R_{\rm FeII}}
    \label{eq9}
\end{equation}

These above analytical forms (Equations \ref{eq5}, \ref{eq6}, \ref{eq8} and \ref{eq9}) are tabulated in Table \ref{tab:table1}. We highlight the resulting values for the product of ionization parameter (\u{}) and local BLR density (\n{}) for the two sources considered in this work, i.e., \textmyfont{NGC 5548} and \textmyfont{I Zw 1}. Since, we later use the SEDs for these two sources, we have the exact value for the $\chi$ reported by {\sc CLOUDY} for them: 0.82 ({\sc NGC 5548}), and, 0.12 ({\sc I Zw 1}). We report the estimates for all the cases accounting for these appropriate $\chi$ values for the two sources in the last two columns in Table \ref{tab:table1}. We will come back to these estimates in Section \ref{sec:filter-EWs-U-n}.

\begin{table}[]
\centering
\caption{Estimates for log(Un$_{\rm H}$) for the various relations considered in this paper}
\label{tab:table1}
\resizebox{\textwidth}{!}{%
\setlength{\tabcolsep}{0.5em} 
\renewcommand{\arraystretch}{2}
\begin{tabular}{c|c|c|c|c|c|c}
\hline
\textbf{Radius-Luminosity relation}  & \textbf{Bolometric Correction} & \textbf{log(Un$_{\rm H}$)}$^{@}$                           & \textbf{NGC 5548}$^{a}$               & \textbf{I Zw 1}$^{b}$   & \textbf{NGC 5548}$^{c}$               & \textbf{I Zw 1}$^{d}$                 \\ \hline
\multirow{2}{*}{\citet{bentz13}} & \citet{richards2006}        & 9.815 - 0.084log$\left(\frac{\rm{L_{5100}}}{10^{44}}\right)$                   & 9.880                     & 9.769 & 10.095                     & 9.149                     \\ \cline{2-7} 
                                     & \citet{netzer2019}                  & 10.050 - 0.284log$\left(\frac{\rm{L_{5100}}}{10^{44}}\right)$                   & 10.272                     & 9.896    & 10.487                     & 9.276                     \\ \hline
\multirow{2}{*}{\citet{du2019}}   & \citet{richards2006}        & 9.625 + 0.1log$\left(\frac{\rm{L_{5100}}}{10^{44}}\right)$ + 0.7R$_{\rm FeII}$  & 9.617  & 10.812 & 9.832  & 10.192 \\ \cline{2-7} 
                                     & \citet{netzer2019}                  & 9.860 - 0.1log$\left(\frac{\rm{L_{5100}}}{10^{44}}\right)$ + 0.7R$_{\rm FeII}$ & 10.008 & 10.939   & 10.223 & 10.319 \\ \hline
\end{tabular}%
}
{\footnotesize
$^{@}$ denotes for the case with $\chi$=0.5 which is used to estimate values for {\sc NGC 5548} and {\sc I Zw 1} in columns 4 and 5, respectively. AGN optical luminosity at 5100\AA\ (L$_{5100}$) for: $^{a}$ \textmyfont{NGC 5548} = 1.66$\times$10$^{43}$ erg s$^{-1}$ \citep{fausnaughetal2016}; $^{b}$ \textmyfont{I Zw 1} = 3.48$\times$10$^{44}$ erg s$^{-1}$ \citep{persson1988}. These are consistent with their respective SEDs considered in this paper. The corresponding \rfe{} for (a) \textmyfont{NGC 5548} = 0.1$\pm$0.02 \citep{du2019}; and for (b) \textmyfont{I Zw 1} is 1.619$\pm$0.060 (Marziani et al. 2021, submitted), respectively. $^{c}$ uses the $\chi$=0.82 as reported by {\sc CLOUDY} for {\sc NGC 5548}, $^{d}$ uses the $\chi$=0.12 as reported by {\sc CLOUDY} for {\sc I Zw 1}, keeping other parameters identical as before.}
\end{table}

\section{Photoionization computations with {\sc CLOUDY}}
\label{sec:cloudy}

We apply the photoionization setup prescription similar to that was demonstrated in \citet[][]{panda2021}. We describe briefly the setup here - we perform a suite of \textmyfont{CLOUDY} \citep[version 17.02,][]{f17} models\footnote{N(\u{}) $\times$ N(\n{}) $\times$ N(Z) = 29$\times$33$\times$3 = 2871 models}
by varying the mean cloud density over a broad range, $10^{5} \leq n_H \leq 10^{13}\;(\rm{in\;cm^{-3}}$), as well as the ionization parameter, $-7 \leq \log U \leq 0$. We consider the gas cloud at a cloud column density, $N_{\rm{H}}$ = $10^{24}$ cm$^{-2}$. We consider two spectral energy distributions (SEDs) - one for \textmyfont{NGC 5548} and the other for \textmyfont{I Zw 1}. We show the SEDs covering the optical-to-X-ray energy range in Figure \ref{fig:sed}. For the \textmyfont{NGC 5548}, we incorporate the SED from \citet{dehghanianetal2019} that is an extension of the SED shown in \citet{mehdipouretal2015}. The SED was prepared using quasi-simultaneous observations taken in 2013-2014 with XMM-Newton, Swift, NuSTAR, INTEGRAL, Chandra, HST, and two ground-based observatories - Wise Observatory and Observatorio Cerro Armazones. We refer the readers to \citet{mehdipouretal2015} for more details on the spectral modelling and continuum extraction over the broad-band energies. On the other hand, the SED for \textmyfont{I Zw 1} is directly derived from the continuum extraction over the near-infrared to ultraviolet range (between $\sim$1000$\rm{\AA}$-1$\rm{\mu}$m) supplemented with the photometric data points in the X-ray region and wavelengths above 2.5 $\mu$m from the previously used SED \citep{panda_cafe1, panda2021}. We combine almost concomitant spectra for this source observed with HST-Faint Object Spectrograph (FOS, \citealt{bechtold_2002}) in the UV that is complemented with data in the optical (obtained using the 2.15m Complejo Astronomico El Leoncito - CASLEO, \citealt{rodriguezardillaetal2002}) and in the NIR (obtained using the 3.2m NASA Infrared Telescope Facility - IRTF, \citealt{riffel2006}). For the continuum points extraction, we automatically identify the emission lines, and select regions in the spectrum free of them to extract these points. A full description of the procedure can be found in an upcoming work (Dias dos Santos et al. in prep.). We consider three cases for the metallicity - at solar composition (\zsun{}), at 3 times solar (3\zsun{}) and at 10 times solar (10\zsun{}) values to model the emission from the low-ionization line emitting region for \textmyfont{I Zw 1}, while we limit ourselves to only solar metallicity case for modelling \textmyfont{NGC 5548}. This assumption for the metal content for the case of \textmyfont{NGC 5548} is made on the basis of our prior results in modelling this source \citep{panda_2021_ngc5548} where the \hb{} and optical \feii{} emission were successfully modelled using {\sc CLOUDY}.

\subsection{Dust sublimation radius prescription}
\label{sec:dust-sub}

Similar to \citet[][]{panda2021}, We incorporate the prescription from \citet{Nenkova2008} to separate the dusty and non-dusty regime in the BLR, which has a form: 
\begin{equation}
\rm{R_{sub}} = 0.4\left(\frac{L_{\rm UV}}{10^{45}}\right)^{0.5},    
\end{equation}
where $\rm{R_{sub}}$ is the sublimation radius (in parsecs) computed from the source luminosity ($L_{\rm UV}$) that is consistent for a characteristic dust temperature.

\begin{table}[]
\centering
\caption{Estimates for dust sublimation radius for \textmyfont{NGC 5548} and \textmyfont{I Zw 1}.}
\label{tab:table2}
\resizebox{\textwidth}{!}{%
\setlength{\tabcolsep}{0.5em} 
\renewcommand{\arraystretch}{2}
\begin{tabular}{c|c|cc|cc}
\hline
\multirow{3}{*}{\textbf{Source}} & \multirow{2}{*}{\textbf{L$_{\rm 5100}$}} & \multicolumn{2}{c|}{\textbf{k$_{\rm bol}$ \citep{richards2006}}} & \multicolumn{2}{c}{\textbf{k$_{\rm bol}$ \citep{netzer2019}}} \\ \cline{3-6} 
 &  & \multicolumn{1}{c|}{\textbf{L$_{\rm bol}$}} & \textbf{R$_{\rm sub}$} & \multicolumn{1}{c|}{\textbf{L$_{\rm bol}$}} & \textbf{R$_{\rm sub}$} \\
 & \textbf{{[}erg s$^{-1}${]}} & \multicolumn{1}{c|}{\textbf{{[}erg s$^{-1}${]}}} & \textbf{{[}pc{]}}$^{(1)}$ & \multicolumn{1}{c|}{\textbf{{[}erg s$^{-1}${]}}} & \textbf{{[}pc{]}} \\ \hline
\textmyfont{NGC 5548} & 1.66$\times$10$^{43}$ & \multicolumn{1}{c|}{1.537$\times$10$^{44}$} & 0.157 & \multicolumn{1}{c|}{3.786$\times$10$^{44}$} & 0.246 \\ \hline
\textmyfont{I Zw 1} & 3.48$\times$10$^{44}$ & \multicolumn{1}{c|}{3.223$\times$10$^{45}$} & 0.718 & \multicolumn{1}{c|}{4.32$\times$10$^{45}$} & 0.83 \\ \hline
\end{tabular}%
}
{\footnotesize
AGN optical luminosity at 5100\AA\ (L$_{5100}$) for \textmyfont{NGC 5548} and \textmyfont{I Zw 1} are taken from \citet{fausnaughetal2016} and \citet{persson1988}, respectively. These values are same as used in Table \ref{tab:table1}. (1) 1 parsec = 3.086$\times$10$^{18}$ cm.}
\end{table}

This is a simplified version of the actual relation which, in addition to the source luminosity term, contains the dependence on the dust sublimation temperature and the dust grain size. We assume a dust temperature $T_{\rm sub}$ = 1500 K, which has been found consistent with the adopted mixture of the silicate and graphite dust grains, and a typical dust grain size, a=0.05 microns. The dependence of the $\rm{R_{sub}}$ on the temperature is quite small - the exponent on the temperature term is -2.8. On the other hand, the dust grain size is a more complex problem, yet the value adopted is fair in reproducing the characteristic dust sublimation radius in our case \citep[see][for more details]{Nenkova2008,honig2019}. The sublimation radius, hence, is estimated using only the integrated optical-UV luminosity for the two representative sources - \textmyfont{NGC 5548} and \textmyfont{I Zw 1}. This optical-UV luminosity is the manifestation for an accretion disk emission and can be used as an approximate for the source's bolometric luminosity. We note that in this case, the ionizing luminosity that leads to the sublimation is almost close to the bolometric luminosity for both the sources considered in this work. The assumed sublimation temperature, 1500 K, corresponds to an average photon energy of 0.0095 Rydberg, or a frequency $\sim$ 14 (in log-scale). This is the lower limit of the SEDs shown in Figure \ref{fig:sed} and used in our computations. Hence, the value for the $\chi$ (ratio of the L$_{\rm ion}$ to the \lbol{}) is set to unity to retrieve the corresponding pairs of ionization parameter (\u{}) and local density (\n{}). Table \ref{tab:table2} provides the estimates for the $\rm{R_{sub}}$ considering the fixed and variable \kbol{} factors. Using these estimates for the \lbol{} and the sublimation radius ($R_{\rm sub}$) and substituting in Equation \ref{eq1}, we get the values for the product of the \u{} and \n{}. This is not be confused with the BLR density as this product (\u{}\n{}) relates to the dust sublimation radius and not the BLR photoionization radius, i.e., \rblr{}. For the 4 pairs of (\lbol{},$R_{\rm sub}$) tabulated in Table \ref{tab:table2} we get value for \u{}\n{}: (a) for \textmyfont{NGC 5548}, 7.9016 (for the fixed \kbol{}) and 7.9030 (for the luminosity-dependent \kbol{}); (b) for \textmyfont{I Zw 1}, 7.9027 (for the fixed \kbol{}) and 7.9040 (for the luminosity-dependent \kbol{}).

\subsection{Estimating the EWs for the low-ionization emission lines in the BLR}
\label{sec:ews_from_cloudy}

\begin{figure}[!htb]
    \centering
    \includegraphics[width=\textwidth]{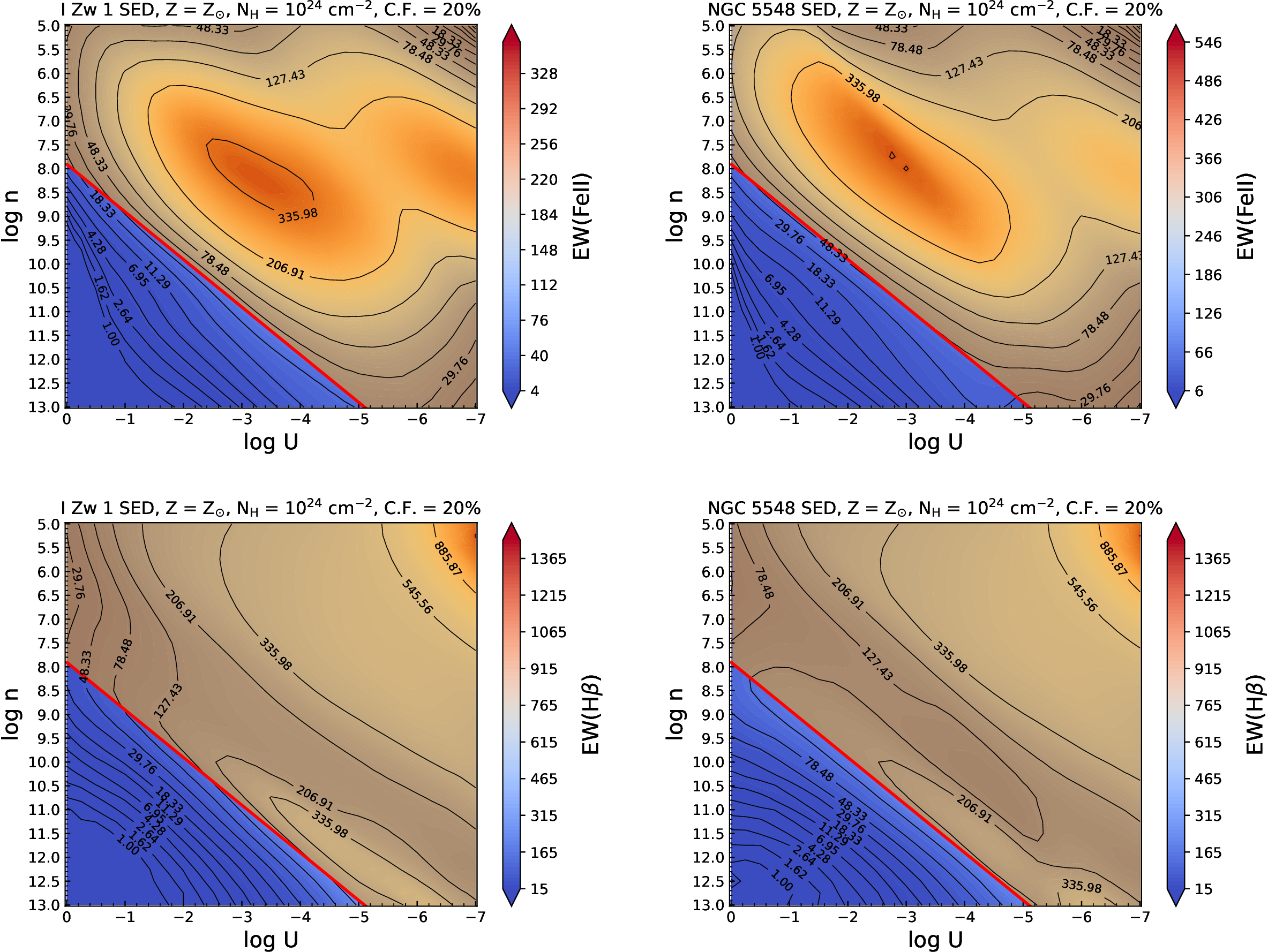}
    \caption{\un{} 2D histograms color-weighted by (top panels) equivalent width (EWs) \feii{}, and (bottom panels) EW(\hb{}). The EWs are computed assuming the continuum luminosity at 5100 \AA\ and at 20\% covering fraction. Two different spectral energy distributions (SEDs) as previously shown in Figure \ref{fig:sed} are used representing (left panels) \textmyfont{I Zw 1} - a prototypical Population A object with high \rfe{}, and (right panels) \textmyfont{NGC 5548} - a prototypical Population B object with relatively low \rfe{}. Models are shown for a characteristic BLR cloud with solar composition (Z=\zsun{}) and column density, N$_{\rm{H}}\,$=$\,10^{24}\,\rm{cm^{-2}}$. The solid red lines in each panel marks the boundary between the non-dusty (in blue) and the dusty (in orange). This is set uniquely for each case of SED using the luminosity-dependent \kbol{} \citep{netzer2019} values for the \lbol{} and $R_{\rm sub}$ as shown in Table \ref{tab:table2}.}
    \label{fig:compare_IZw1_NGC5548_2dhist_EW_20pc}
\end{figure}

To estimate the EWs for \hb{} and optical \feii{}, we use the continuum luminosity given by {\sc CLOUDY} for each model as a reference. By default the EWs extracted with this approach assumes 100\% covering factor. We then re-scale this value to 20\% of its original value. The assumption of 20\% has been shown to be reliable adhoc estimate for the covering factor \citep{korista_goad_2000, baldwin2004, sarkar2020, panda2021}. In Figure \ref{fig:compare_IZw1_NGC5548_2dhist_EW_20pc}, we illustrate the result for the two cases of SEDs (\textmyfont{I Zw 1} and \textmyfont{NGC 5548}) with the base setup, i.e., at solar metallicity (\zsun{}) and cloud column density, N$_{\rm{H}}\,$=$\,10^{24}\,\rm{cm^{-2}}$. The upper and lower panels show the \un{} parameter space with the auxiliary axis (colormap) depicting the EW(\feii{}) and the corresponding EW(\hb{}), respectively. The threshold for the dusty (shaded in orange) and dustless (in blue) line emitting region is set using the prescription described in the previous section (Sec. \ref{sec:dust-sub}). We use the luminosity-dependent \kbol{} versions of the \u{}\n{} (in log-scale), i.e., 7.9040 for the \textmyfont{I Zw 1}, and 7.9030 for the \textmyfont{NGC 5548}. Henceforth, we will only discuss the results in the context of emission from the dustless BLR.

\begin{figure}[!htb]
    \centering
    \includegraphics[width=\textwidth]{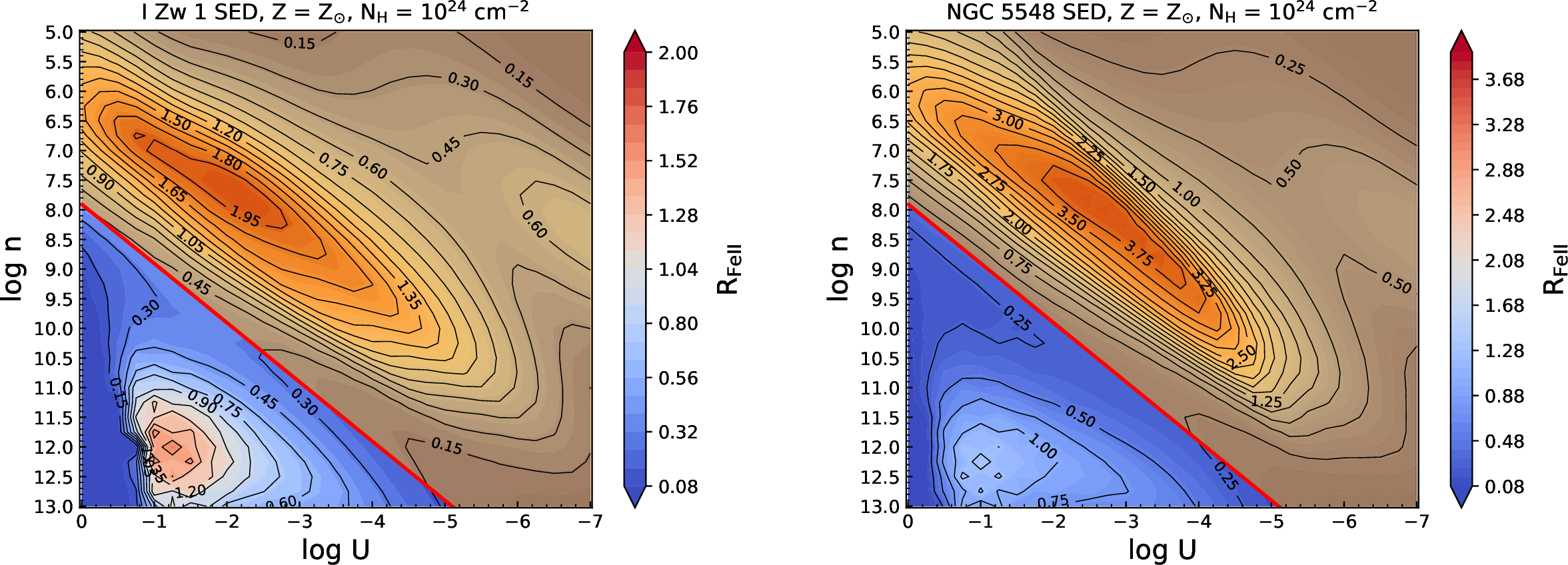}
    \caption{\un{} 2D histograms color-weighted by \rfe{}. Remaining labels and parameters shown here are identical to Figure \ref{fig:compare_IZw1_NGC5548_2dhist_EW_20pc}.}
    \label{fig:compare_IZw1_NGC5548_2dhist_rfe}
\end{figure}

In Figure \ref{fig:compare_IZw1_NGC5548_2dhist_rfe}, we show the \un{} parameter space with the auxiliary axis depicting the ratio \rfe{}, i.e. the EW(\feii{}) to the EW(\hb{}) for the two SEDs. We have assumed that the two emission lines are produced from a similar region in the BLR \citep[see e.g.,][]{barth_2013,hu15,panda18b,gaskell_2021_FeII} and hence the covering factor is set to be equal for both the emission lines.

We also consider a case with a higher covering factor (i.e., 60\%) to highlight the effect due to non-radial motions \citep{kollatschny_zetzl_2013} or changes in the accretion disk structure \citep{abramowicz88, Wang2014}. This higher value for the covering factor is an upper threshold as modelled in the locally optimized cloud models by \citet{korista_goad_2000} for \textmyfont{NGC 5548}. This is shown in Figure \ref{fig:compare_IZw1_NGC5548_2dhist_EW_60pc} under the same conditions as for the case with the 20\% covering factor.

\subsection{Comparison with reverberation-mapped estimates}
\label{sec:compare-w-data}

\begin{figure}[!htb]
    \centering
    \includegraphics[width=\textwidth]{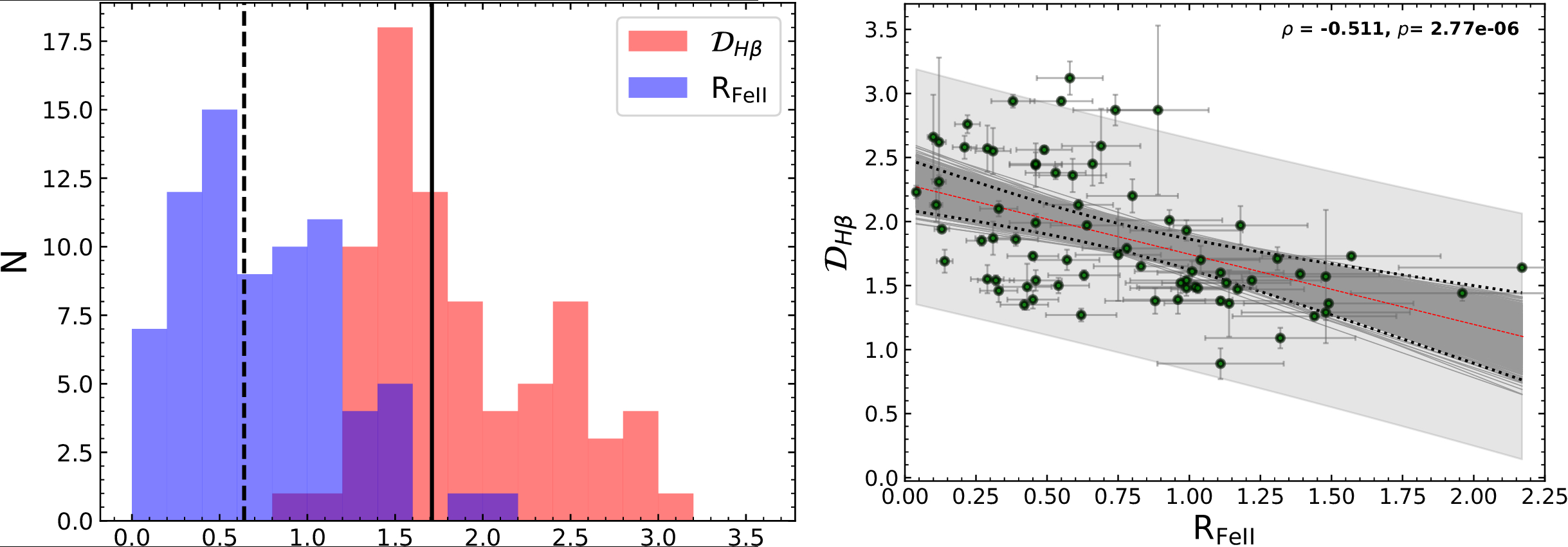}
    \caption{(Left:) Distribution of the $\mathcal{D_{\rm H\beta}}$ (in red) and \rfe{} (in blue) for the sample from \citet{du2019} and as shown in Table \ref{tab:rm-data}. The median values for the two parameters are shown using the vertical lines - solid ($\mathcal{D_{\rm H\beta}}$) and dashed (\rfe{}). (Right:) Correlation between the $\mathcal{D_{\rm H\beta}}$ and \rfe{} for this sample (see also Figure 2 in \citet{du2019}). The Spearman's rank correlation coefficients ($\rho$) and the $p-$values are also reported. The ordinary least-square (OLS) fit for each panel is shown using red solid line. Black dotted lines mark the confidence intervals at 95\% for the 1000 realizations (dark gray lines) of the bootstrap analysis. The corresponding prediction intervals are shown in the background using light gray color.}
    \label{fig:D-RFe-histogram-Du2019}
\end{figure}

To make a quantitative comparison with the results from the {\sc CLOUDY} simulations, we utilize the sample of reverberation-mapped AGNs from \citet{du2019}. The sample consists of 75 AGNs for which an independent and homogeneous spectral fitting in the optical region (including the 4430–5550\AA~ window in the rest frame) was performed in their paper. The spectral window includes the \hb{} and optical \feii{} emission blend (between 4434-4684 \AA) that is necessary to estimate the ratio, \rfe{}, the bolometric luminosity using the AGN luminosity at 5100\AA~ and the black hole mass using the FWHM(\hb{}) in association with the distance of the BLR from the central continuum source which is obtained from various reverberation mapping campaigns \citep[see Section 2.1 in ][where they list the various campaigns]{du2019}, and thus the Eddington ratio or its equivalent - the dimensionless accretion rate ($\mathcal{\dot{M}}$). We present the spectral and derived properties of this sample in Table \ref{tab:rm-data}, that includes the AGN luminosity at 5100\AA~ (\l5100{}), the black hole mass (\mbh{}), the $\mathcal{\dot{M}}$, \rfe{} and the $\mathcal{D_{H\beta}}$, where the latter is the ratio of the FWHM(\hb{}) to the dispersion of the \hb{}. To estimate the $\mathcal{\dot{M}}$, the authors \citep[e.g., ][]{wangetal13,du2015,dupu2016L,du2019} use the following form:

\begin{equation}
    \mathcal{\dot{M}} = 20.1\left(\frac{l_{44}}{\rm{cos}\; i}\right)^{\frac{3}{2}}m_{7}^{-2}
\end{equation}

where, the $l_{44}$ is the AGN luminosity at 5100\AA~ in the units of 10$^{44}$ erg s$^{-1}$, the $m_{7}$ is the black hole mass in the units of 10$^7$ \msun{}, and $i$ is the inclination angle of the accretion disk. The estimates for the \mdot{} tabulated in their paper \citep{du2019} and in Table \ref{tab:rm-data} assume an average value of the cos $i$ = 0.75. Since the \mdot{} and Eddington ratio are equivalent, we can express the Eddington ratio (\LLEdd{}) as follows:

\begin{equation}
    \frac{L_{\rm bol}}{L_{\rm Edd}} = 7.455\times 10^{-18} \left(k_{\rm bol} M_{\rm BH}^{\frac{1}{3}} \mathcal{\dot{M}}^{\frac{2}{3}}\right)
\end{equation}

Thus, any inferences that will be drawn based on the \mdot{} can be extended directly to the corresponding estimates of the Eddington ratios. We decided to use this particular sample to ensure that the sources are treated in the same manner. This removes the bias from fitting techniques employed by different groups. An independent spectral fitting incorporating newer measurements and newer sources is needed which is outside the scope of this paper.

Next, we show the distribution of the estimates for the \rfe{} and for the $\mathcal{D_{H\beta}}$ for this sample in the left panel of Figure \ref{fig:D-RFe-histogram-Du2019}. The range of \rfe{} values lie between [0.04, 2.17], while for the $\mathcal{D_{H\beta}}$ this range is between [0.89, 3.12]. Thus, according to the definition of the spectral sub-types in the optical plane of the main sequence \citep[see Figure \ref{fig:sed}, also in][]{mar18,panda19b,panda19c}, this sample is dominated by Population A sources (51/75) and covers the spectral types from A1-A4. Here A1 are the more typical, low-\rfe{} ($\leq$ 0.5) AGNs and the A4 are the more rare, strong-\feii{} emitters ($\geq$ 2.0). In addition, we have a considerable number of Population B sources in this sample (24/75) including the \textmyfont{NGC 5548}. We refer the readers to Table 1 in \citet{du2019} for the estimates for the FWHMs for the sources in the sample. In the right panel of \ref{fig:D-RFe-histogram-Du2019}, we show the strong anti-correlation between \rfe{} and $\mathcal{D_{H\beta}}$ (see also Figure 2 in \citealt{du2019}). This figure re-iterates an already known fact, i.e., the sources with high \rfe{} (especially $\gtrsim$ 1) also are found to have higher accretion rates and this then affects the emission line profiles - making them more Lorentzian, as opposed to the generally well-fitted Gaussian profiles that is suited for the sources with low \rfe{} estimates (e.g. typical Population B sources). This change in the line profiles affects directly the value of the $\mathcal{D_{H\beta}}$ - for a pure Lorentzian this value tends to zero, while for a single Gaussian this value is 2$\sqrt{2 \rm{ln}(2)}$ = 2.35. For a rectangular function, $\mathcal{D_{H\beta}}$ = 2$\sqrt{3}$ = 3.46 \citep{collin2006}. We describe the analytical formulations including the $\mathcal{D_{\rm H\beta}}$ in our prescription in the Appendix \ref{app:D-parameter-paramterization}.

\section{Results}
\label{sec:results}

\subsection{Understanding the \un{} parameter space for {\sc I Zw 1} and {\sc NGC 5548}}
\label{sec:un-rfe-Izw1-NGC5548}

In the following sections, we describe the results from the various {\sc CLOUDY} photoionization models that were made to constrain the physical parameter space in terms of the \un{} diagrams. The key here is the focus on the recovery of the EWs for the two low ionization emission lines - \hb{} and \feii{}, in addition to the ratio of their EWs, i.e., \rfe{}. In Figure \ref{fig:compare_IZw1_NGC5548_2dhist_EW_20pc}, we depict the EW(\feii{}) and the EW(\hb{}) for the two sources (\textmyfont{I Zw 1} and \textmyfont{NGC 5548}) under the assumption that the BLR in the two cases has solar composition and the covering factor is identical, i.e. 20\%. Another important highlight is the separation of the dustless region from the region where dust can survive. Species like the \feii{} get strongly depleted in the presence of dust and can be used as a tracer for the dust in the extended, intermediate-line regions that are located further away from the BLR \citep[see e.g.,][]{adhikari16}. As described in Section \ref{sec:dust-sub}, we have made a simple assumption on the location of the dust sublimation radius that is effectively dependent only on the AGN luminosity. This uniquely sets the dust sublimation radius for each source (see Table \ref{tab:table2}). In Figure \ref{fig:compare_IZw1_NGC5548_2dhist_EW_20pc} (and henceforth), we have used the dust sublimation radius case assuming the luminosity-dependent \kbol{} correction that gives a slightly larger value for this radius. In the figure, the radius ($R$) is shown using a red solid line which corresponds not to the \rblr{} but to a radius that is much larger than \rblr{}. The values for this larger $R$ in terms of \u{}\n{} are very similar for the two sources as the differences in their luminosities and radial extensions almost balances out -  (a) for \textmyfont{NGC 5548}, 7.9030 (for the luminosity-dependent \kbol{}); (b) for \textmyfont{I Zw 1}, 7.9040 (for the luminosity-dependent \kbol{}). The corresponding \rfe{} estimates for the two cases (see Figure \ref{fig:compare_IZw1_NGC5548_2dhist_rfe}) also have similar demarcations.

\begin{figure}[!htb]
    \centering
    \includegraphics[width=\textwidth]{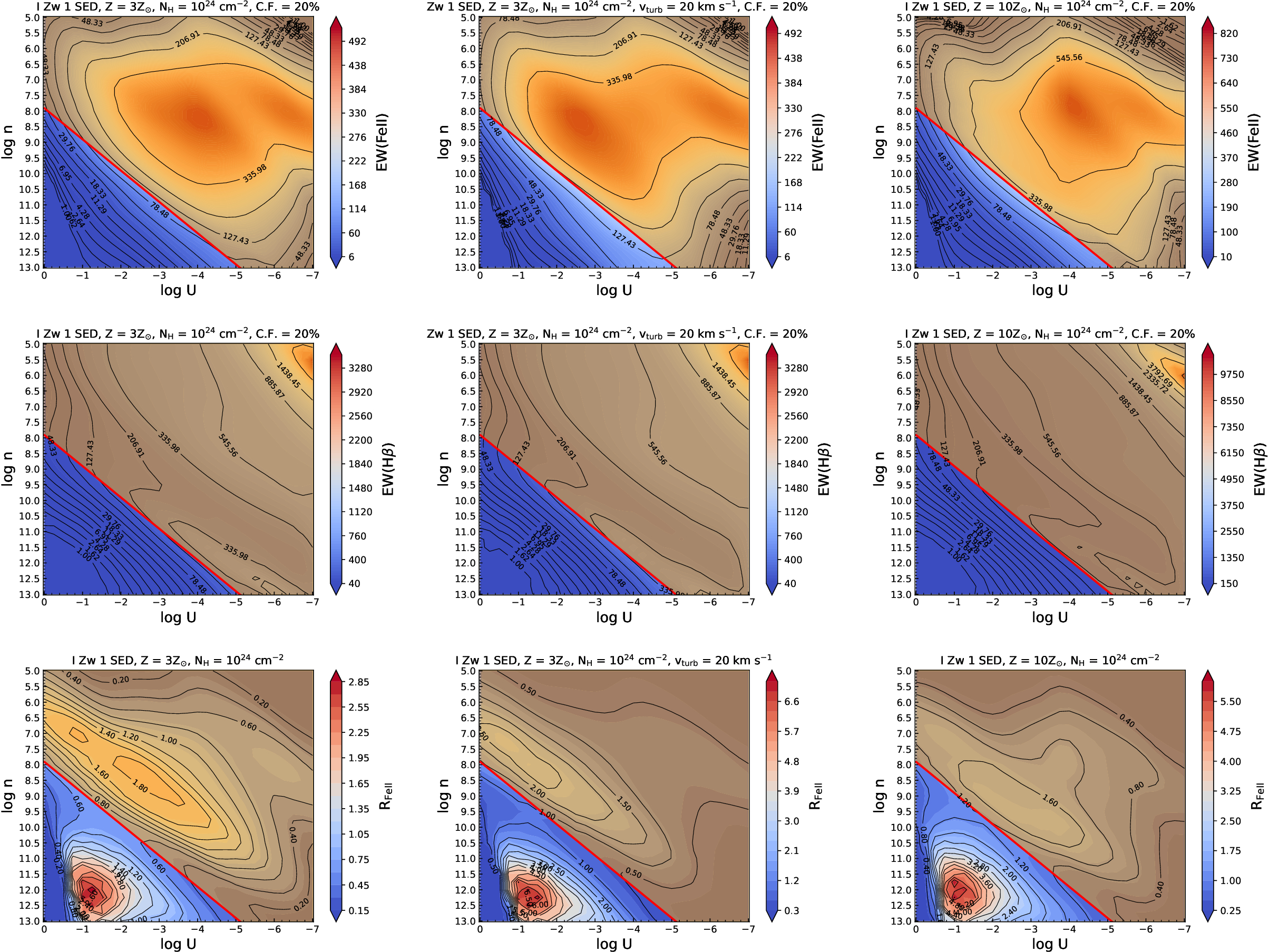}
    \caption{\un{} 2D histograms color-weighted by (top panels) equivalent width (EWs) \feii{}, (middle panels) EW(\hb{}), and (bottom panels) \rfe{}. The labels and parameters shown here are identical to Figure \ref{fig:compare_IZw1_NGC5548_2dhist_EW_20pc}. The panels depict the models considered for a BLR metallicity of 3\zsun{} (left panels), then for a metallicity of 3\zsun{} with additional microturbulence of 20 \kms{} (middle panels), and, for a metallicity of 10\zsun{} (right panels).}
    \label{fig:compare_IZw1_3Z-3Zv20-10Z}
\end{figure}

\subsection{Comparing the reverberation-mapped sources with the {\sc CLOUDY} models}

Another way of looking at this scenario is by comparing the product of the \u{} and \n{} directly versus the \rfe{}. This is already shown from the estimates tabulated in our Table \ref{tab:table1} for the two sources. But, as there are the various considerations for the \kbol{} and the \RL{} relations, the values obtained for the photoionization radius estimator, i.e., the product \u{}\n{}, varies, albeit slightly. In this section, we organize the \un{} parameter space from each model and compare the relevance of these results to the reverberation-mapped sources with spectral coverage that includes the \rfe{} measurements. We intend to assess the changes in the SED of the two prototypical sources considered in work to see if they account for the \rfe{} estimates reported from spectral fitting. We highlight the salient differences between the two cases, and how our analytical prescriptions reported in Section \ref{sec:analytical_description} perform against the numerical estimates from {\sc CLOUDY}.

\begin{figure}[!htb]
    \centering
    \includegraphics{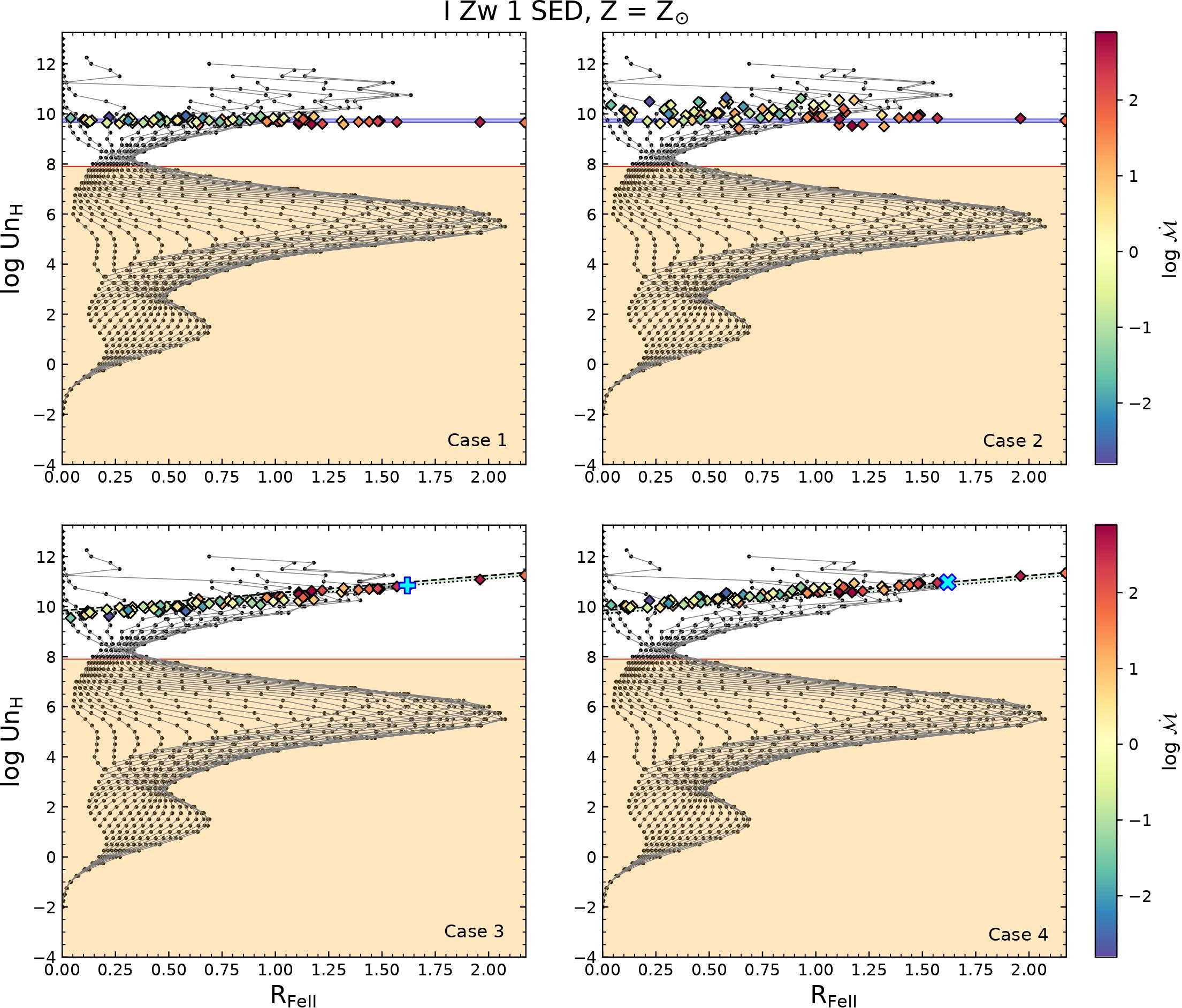}
    \caption{Grid composed of the family of distributions for the \u{}\n{} (in log-scale) as a function of the \rfe{} (gray lines with black dots) from one of our {\sc CLOUDY} simulations - \textmyfont{I Zw 1} SED, $N_{\rm{H}}$ = $10^{24}$ cm$^{-2}$, and, at solar composition. The shaded region in orange depicts the region within the dust and the solid red line marks the location of the dust sublimation radius as per Table \ref{tab:table2}. The sample of 75 reverberation-mapped sources (see Table \ref{tab:rm-data}) are shown for the corresponding cases of \u{}\n{} as per the four analytical forms shown in Table \ref{tab:table1}. These sources are color-coded with their corresponding \mdot{} estimates. The location of the source \textmyfont{I Zw 1} is shown using a blue cross for the lower panels where there is an explicit dependence on the \rfe{}.}
    \label{fig:compare-IZw1-4cases}
\end{figure}

In Figure \ref{fig:compare-IZw1-4cases}, we show the grid from one of our {\sc CLOUDY} simulations (\textmyfont{I Zw 1} SED, $N_{\rm{H}}$ = $10^{24}$ cm$^{-2}$, and, at solar composition). The grid is composed of the family of distributions for the \u{}\n{} (in log-scale) as a function of the \rfe{}. These are identical to what we show in the left panel of Figure \ref{fig:compare_IZw1_NGC5548_2dhist_rfe} just represented differently. The shaded region in orange depicts the region within the dust and the solid red line marks the location of the dust sublimation radius as per Table \ref{tab:table2}. The location of the dust sublimation radius is neatly poised at one of the minima for the \rfe{}.

Preparing broad-band SEDs for the diverse population of AGNs is not easy, especially to get contemporaneous spectral or photometric data over a wide range of energies. Having SEDs that can be representative of the sub-populations, e.g., Population A and Population B, is quite useful. This was our intention from this work. To test the validity of our models on the observational estimates for sources with spectral coverage and reverberation mapping, we overlay the 75 sources from Table \ref{tab:rm-data} on these maps. These sources are colour-coded as a function of the dimensionless accretion rate (\mdot{}). For each of the four panels, we evaluate the \u{}\n{} for each source using the four relations as shown in the Table \ref{tab:table1}, i.e., accounting for (a) the standard \RL{} relation from \citet{bentz13} with fixed \kbol{} (Case 1) and with luminosity-dependent \kbol{} (Case 2); and (b) the \rfe{}-dependent \RL{} relation from \citet{du2019} with fixed \kbol{} (Case 3) and with luminosity-dependent \kbol{} (Case 4). For the first two cases (upper panels), we show the \u{}\n{} for \textmyfont{I Zw 1} using the \l5100{} as also reported in Table \ref{tab:table1}. These are shown using the two solid blue lines on the upper panels. As we can notice, the location of the sources on this plane is affected due to the change of the \kbol{} - a larger scatter in the \u{}\n{} is seen especially for sources that have relatively low \rfe{} and low to mid \mdot{}. For the latter cases (lower panels), we show the \u{}\n{} for \textmyfont{I Zw 1} which is now dependent on the \l5100{} and \rfe{}, and hence, changes with the change in the \rfe{}. These are shown using the dashed (fixed \kbol{} case) and dotted (luminosity-dependent \kbol{}) black lines on the panels. The location of \textmyfont{I Zw 1} is shown using the blue plus and a blue cross symbol, respectively, in these two panels. The scatter in these panels is lower compared to Case 2 (upper right panel). The \textmyfont{I Zw 1} SED under the solar composition can incorporate a large fraction of the sources in the sample, although as expected the observed \rfe{} estimates for \textmyfont{I Zw 1} and two other sources (SDSS J101000 and IRAS 04416+1215) is higher than predicted from these base models. 

\begin{figure}[!htb]
    \centering
    \includegraphics{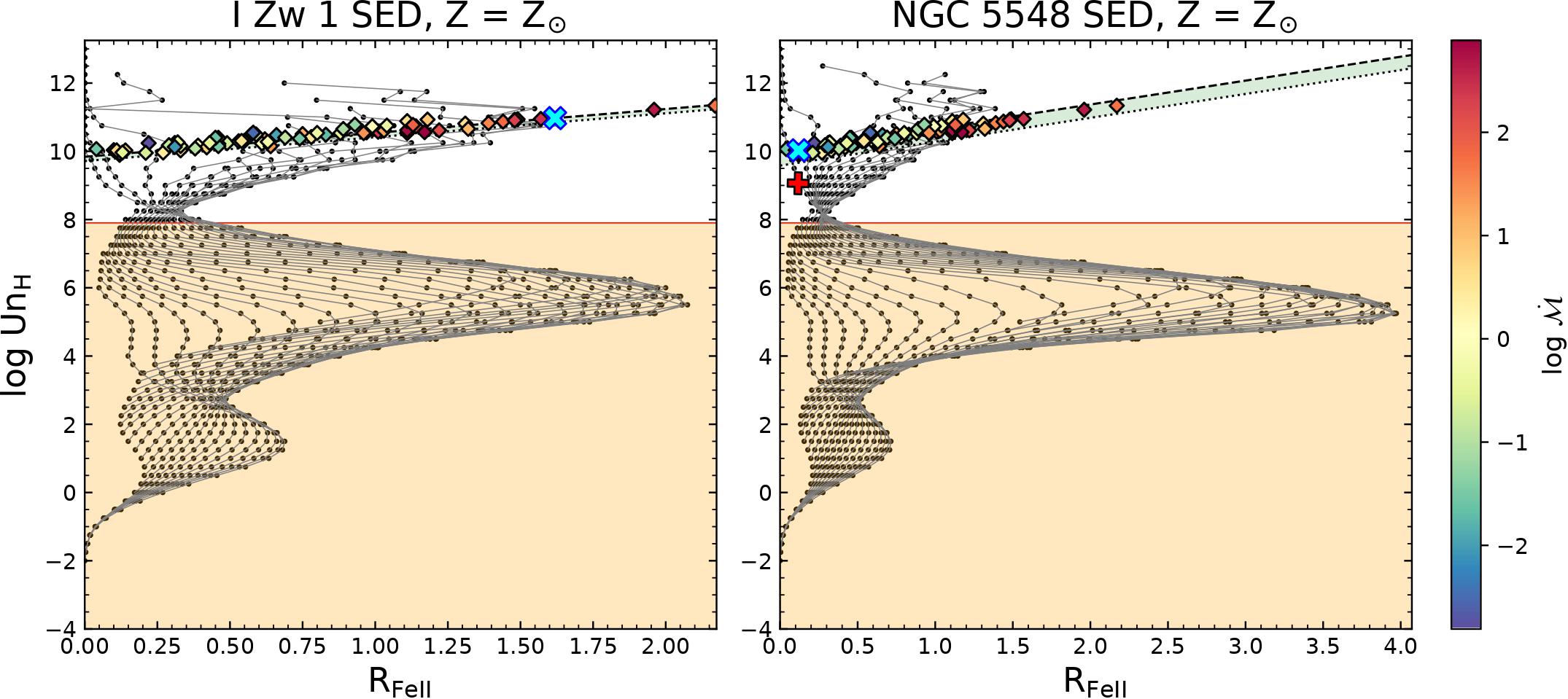}
    \caption{Similar to the last panel (Case 4) in Figure \ref{fig:compare-IZw1-4cases}. Here, we compare the solar composition model for the two sources - \textmyfont{I Zw 1} and \textmyfont{NGC 5548}. The location of the respective sources are marked using a blue cross in the corresponding panels. The red plus symbol on the right panel marks the position of the value obtained for \u{}\n{} using the filtering of EWs (see Section \ref{sec:filter-EWs-U-n}).}
    \label{fig:grid_IZw1_NGC5548_wo_D}
\end{figure}

For completeness, we also show a comparison between the two SED cases at solar composition with the observed sample side-by-side in Figure \ref{fig:grid_IZw1_NGC5548_wo_D}. The left panel is identical to Case 4 already shown in Figure \ref{fig:compare-IZw1-4cases}. The right panel shows the case with the \u{}\n{} grid extracted from our {\sc CLOUDY} simulations but for the \textmyfont{NGC 5548} SED. Like in the bottom panels of Figure \ref{fig:compare-IZw1-4cases}, we show the \u{}\n{} from the analytical relations for the two panels dependent on the \l5100{} and \rfe{} (shown using the dashed (fixed \kbol{} case) and dotted (luminosity-dependent \kbol{}) black lines on the panels). We also locate the sources in the corresponding panels using a blue cross symbol. We can notice that the \textmyfont{NGC 5548} sits among the lowest \rfe{} sources. It agrees well in all of the four cases shown earlier in Figure \ref{fig:compare-IZw1-4cases} and hence is rather unaffected by the inclusion/exclusion of the fixed/variable \kbol{} or the change in the \RL{} relation used to infer the \u{}\n{}. Right away we notice that the \textmyfont{NGC 5548} case predictions encompass a lower fraction of the sources than \textmyfont{I Zw 1} case, especially those with reportedly high \rfe{} values. The strip showing the \u{}\n{} from the Case 3 and Case 4 is thicker in the \textmyfont{NGC 5548} that is due to the effect of luminosity (see Table \ref{tab:table1} with the \RL{} relation cases with \rfe{}-dependence). The location of the observed sources is the same in these two panels, but the extent of the overall \rfe{} predicted with the \textmyfont{NGC 5548} case from the models is higher - the dominant peak is located well within the dusty region. Thus, the \textmyfont{I Zw 1} case predicts higher \rfe{} in the dustless region, as expected from the observed values. We also mark a red plus symbol on the right panel of this figure which marks the position of the value obtained for \u{}\n{} after careful filtering of the possible pairs of solutions by comparing the EWs of both the \feii{} and the \hb{} belonging to the non-dusty part of the BLR. We expand more on this issue in the next section (see Section \ref{sec:filter-EWs-U-n}). We also make a comparative analysis between the two cases including the $\mathcal{D_{\rm H\beta}}$ parameterization within our analytical formalism (see Figure \ref{fig:grid_IZw1_NGC5548_w_D}). 

\begin{figure}[!htb]
    \centering
    \includegraphics{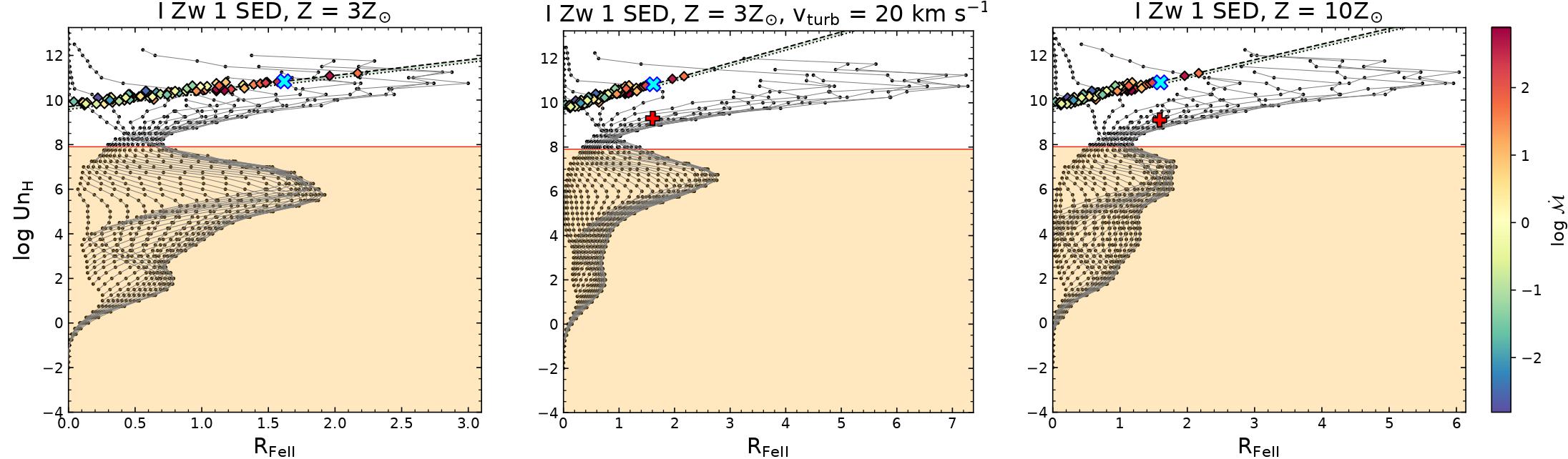}
    \caption{Similar to the last panel (Case 4) in Figure \ref{fig:compare-IZw1-4cases}. Here, we compare \textmyfont{I Zw 1} models at 3\zsun{}, at 3\zsun{} with an additional microturbulent velocity 20 \kms{}, and at 10\zsun{}. The location of the respective sources are marked using a blue cross in the corresponding panels. The red plus symbol on the right panel marks the position of the value obtained for \u{}\n{} using the filtering of EWs (see Section \ref{sec:filter-EWs-U-n}).}
    \label{fig:un-grids-3Z-3Zv20-10Z}
\end{figure}

Subsequently, following the results that were obtained earlier with the \textmyfont{I Zw 1} necessitating an increased metal content (and turbulent motions) within the BLR (see Section \ref{sec:un-rfe-Izw1-NGC5548}), we show the grids of \u{}\n{} for each of the three cases - at 3\zsun{}, at 3\zsun{} with 20 \kms{} microturbulence, and finally, the case with the 10\zsun{}. Figure \ref{fig:un-grids-3Z-3Zv20-10Z} shows these three cases. The dominant peak in these cases shift to the region that corresponds to the dustless BLR (i.e., with \u{}\n{} $\gtrsim$ 8.0). Already in just the 3\zsun{} case (left panel in Figure \ref{fig:un-grids-3Z-3Zv20-10Z}), all the observed estimates are well within the grid lines extracted from the models. But as we have emphasized before, this needs to be supplemented with the EWs recovered for the sources using the models. With these plots, we wanted to show that the effect of the SED with the added contribution of the metal content and microturbulence can significantly affect the recovery of the \rfe{} and that SEDs for prototypical sources (like \textmyfont{NGC 5548} and \textmyfont{I Zw 1}) can be used to infer properties of sources alike. Similar to the previous figure (the right panel with \textmyfont{NGC 5548}), we show the value obtained for \u{}\n{} after the EW-filtering using a red plus symbol.

\subsection{Bringing it all together - filtering the optimal (\u{},\n{}) and EWs}
\label{sec:filter-EWs-U-n}
Now focusing our attention to the non-dusty part of the BLR, we would like to compare the estimates for the EWs for these two lines obtained from the observations and extract the optimal pairs of solution(s) for the \u{} and \n{}. We use the estimates that were quoted by \citet{du2019} for (a) \textmyfont{NGC 5548}: EW(\feii{}) = 11.8$\pm$1.4, EW(\hb{}) = 117.8$\pm$27.3 (this gives an \rfe{}=0.1$\pm$0.02); and from Marziani et al. (2021, submitted) for (b) \textmyfont{I Zw 1}: EW(\feii{}) = 72.86$\pm$15.04, EW(\hb{}) = 45.0$\pm$9.42 (this gives an \rfe{}=1.619$\pm$0.06).

We can see from the right panels in Figure \ref{fig:compare_IZw1_NGC5548_2dhist_EW_20pc} which is depicting the case of \textmyfont{NGC 5548}, that we are successful in recovering these estimates for the EWs for the two species. Although, for the case of \textmyfont{I Zw 1}, while we are almost able to get the match for the EW(\hb{}), the EW(\feii{}) is considerably underestimated. The higher values for the EWs are seen for the regions which are shrouded in the dust where both the ionization parameter and BLR densities are quite low from the viewpoint of the BLR emission \citep[see e.g.,][]{negreteetal12,mar18,panda18b,sniegowska+20}. There is a slight increase when we consider a higher covering factor (60\%, see Figure \ref{fig:compare_IZw1_NGC5548_2dhist_EW_60pc}) but still a deficit of $\sim$10-20\AA~ is found for this case. Next, we proceed on to testing with higher metal content in the BLR. In Figure \ref{fig:compare_IZw1_3Z-3Zv20-10Z}, we show the results for the consideration of two cases of super-solar metallicity - 3\zsun{} (left panels) and 10\zsun{} (right panels). We have again assumed the 20\% covering factor to estimate the EWs. As we can notice, the case with 3\zsun{} is still unable to recover the EW(\feii{}) as suggested from the observations. But, when we go to an even higher metal content (10\zsun{}), we are eventually successful. On the other hand, the BLR cloud can be locally turbulent \citep{baldwin2004, bruhweiler08, shields2010, kollatschny_zetzl_2013} and it has been shown to substantially affect the \feii{} spectrum by facilitating continuum and line-line fluorescence \citep[see e.g.,][]{shields2010,panda18b,panda19,sarkar2020}. We consider a microturbulence value of 20 \kms{} suggested by our previous works \citep{panda18b,panda19} and complement it with the case at 3\zsun{}. The middle panels in Figure \ref{fig:compare_IZw1_3Z-3Zv20-10Z} show the results from this model. We notice that this case can reproduce the EW(\feii{}) as well, in addition to the successful recovery of the EW(\hb{}) and hence, the \rfe{}. \textit{We would like to emphasize that the suggested solutions in terms of \u{} and \n{} are not the ones that show the maximum \rfe{}, rather the ones where both the EWs and the \rfe{} are in agreement with the observations}. Thus, a small microturbulence can affect the recovery of the \feii{} and hence the \rfe{} and the model thus don't necessitate the exceptionally high metal content. For \textmyfont{I Zw 1}, we find that the best agreement is obtained with a metal content that is slightly super-solar (Z$\gtrsim$3\zsun{}) with the inclusion of turbulent motions within the BLR cloud \citep[see also][for an overview on the effect of microturbulence in \textmyfont{I Zw 1}]{panda2021}.

\begin{figure}
    \centering
    \includegraphics[width=\textwidth]{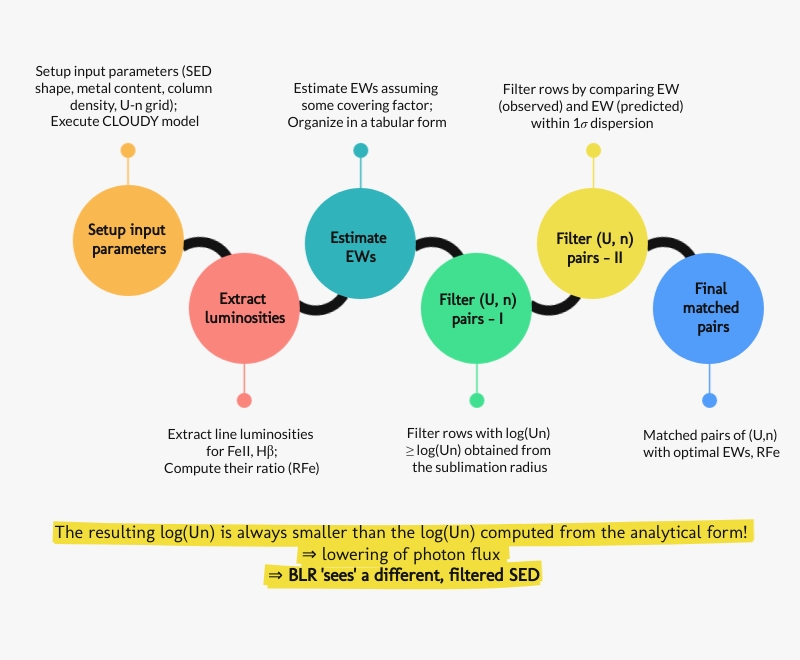}
    \caption{Flowchart depicting the steps taken in the filtering process - starting from the preparation of the input setup till procuring the matched pairs of (\u{},\n{}) for each model described in this paper.}
    \label{fig:flowchart}
\end{figure}

\begin{figure}
    \centering
    \includegraphics[width=\textwidth]{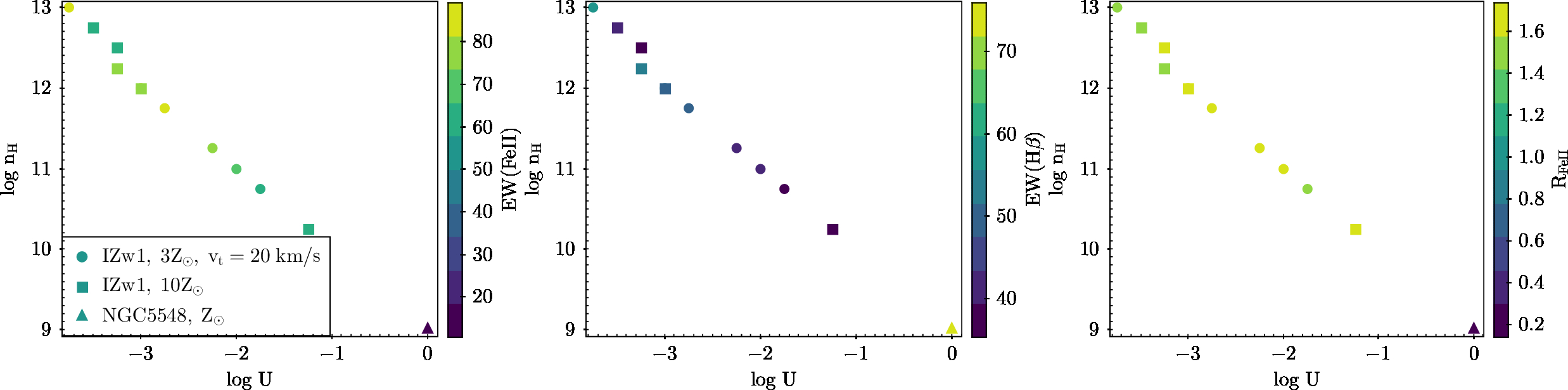}
    \caption{The \un{} parameter space as a function of: (a) EW(\feii{}), (b) EW(\hb{}), and, (c) \rfe{}. These mark the final filtered solutions adopting the steps shown in Figure \ref{fig:flowchart}. The three successful cases (two for \textmyfont{I Zw 1}, and, one for \textmyfont{NGC 5548}) are shown using respective symbols.}
    \label{fig:filtered-solutions}
\end{figure}

In order to finalize the pairs of (\u{},\n{}) we illustrate our filtering process in Figure \ref{fig:flowchart}. This outlines how we arrive at the solutions for these physical parameters in the non-dusty part of the BLR that best represent the conditions in the \hb{} and \feii{} line emitting region. We start with making a first filtering by accounting for the subset of the \un{} where their product is at or larger than the \u{}\n{} estimated using the dust sublimation radius for the corresponding cases, e.g. for \textmyfont{I Zw 1} considering the luminosity-dependent \kbol{}, we have the \u{}\n{} (in log-scale) = 7.9040. Next, we filter from the remaining set of those that agree in their EWs (simultaneously for \feii{} and \hb{}) predicted by {\sc CLOUDY} to their observed values within 1-$\sigma$ dispersion (of the observed value). The final remaining solutions are plotted in Figure \ref{fig:filtered-solutions} for the cases where the agreement is found on all counts. We see that as we expected, for \textmyfont{I Zw 1}, the cases with 3\zsun{} with a small microturbulent velocity (v$_{\rm turb}$ = 20 \kms{}), and the case with 10\zsun{}, are well-suited. While for \textmyfont{NGC 5548}, the BLR cloud with solar abundance is sufficient. Although in this case we recover only one pair of solution (see the triangle marked in Figure \ref{fig:filtered-solutions}) where the predicted density is quite low and the ionization parameter is significantly higher. The single cloud assumption that we make here to perform the {\sc CLOUDY} modelling needs to be revisited and compared against its counterpart, e.g. the locally optimized cloud model \citep[LOC,][]{baldwin1995,korista_goad_2000}, wherein the setup assumes a system of clouds with distribution in density and location from the central source. The LOC model has been shown to agree better particularly in the case of \textmyfont{NGC 5548}. A subsequent work is under progress that deals with this exact issue. While, in the case of \textmyfont{I Zw 1}, as also discussed in \citet{panda2021}, the increase in the accretion rate may puff up the very inner regions of the accretion disk leading to the BLR receiving a filtered SED, one that is significantly different from the SED that is observed by a distant observer, and with much less ionizing photons. This may further lead to the inward shift in the location of the clouds due to reduced radiation pressure in the BLR region. This may lead to cloud coagulation one that is well-described by a single cloud model.

In Figure \ref{fig:filtered-solutions}, we can get the exact values of the product of the \u{}\n{} which is in the range 9-9.25 for \textmyfont{I Zw 1}, and 9 for \textmyfont{NGC 5548}. Now coming back to the red plus symbol that were marked in the panels of Figures \ref{fig:grid_IZw1_NGC5548_wo_D} and \ref{fig:un-grids-3Z-3Zv20-10Z}, we see that the solution obtained from our analytical formulation (see Table \ref{tab:table1}) and the solution obtained from this filtering process differ by almost 2 dex, i.e., for the luminosity-dependent \kbol{} case with \rfe{}-based \RL{} relation, we get a \u{}\n{} = 10.939. Taking a ratio of the two \u{}\n{} from these different formalisms, we get a value between 1-2\%. This is what is the fraction of the actual number of ionizing photon flux that is received at the BLR that leads to the line-formation and emission of the \hb{} and \feii{} in \textmyfont{I Zw 1}. This is exactly what we realized in our previous work \citep{panda2021} - that the BLR ``sees'' a different, filtered SED with only a very small fraction ($\sim$1\%) that leads to the line emission in the low-ionization emitting region of the BLR. With a rigorous filtering approach, we have confirmed this hypothesis in this work. Similarly for {\sc NGC 5548}, for the luminosity-dependent \kbol{} case with \rfe{}-based \RL{} relation, we get a \u{}\n{} = 10.008. The fraction of the photon flux recovered is $\sim$10\%. We note that these estimates for the fraction of ionizing continuum received by the BLR are obtained for a pre-assumed value of $\chi$ = 0.5 (where $\chi$ is the ratio of L$_{\rm ion}$ to the \lbol{} predicted by CLOUDY for each input SED). Changing the $\chi$ to a value consistent for the {\sc I Zw 1} SED, i.e., 0.12, we get the actual ionizing continuum received by the BLR to be between $\sim$5-10\%. For {\sc NGC 5548}, this fraction is much higher ($\chi$=0.82). Thus, the actual ionizing continuum in this case received by the BLR is still $\sim$10\%.

Hence, through this analysis we realize the importance of the actual ionizing luminosity, in addition to the \l5100{} and \rfe{} estimated from their respective spectra, that recovers the pairs of ionization parameter and local BLR density, one that is representative of the properties of the low-ionization line emitting BLR. This ionizing luminosity is estimated with the knowledge of the exact shape of the SED for the corresponding source. In view of a series of work \citep{negrete_etal_2013,ms14,panda19b,ferlandetal2020,2021Univ....7..484M} that have highlighted the importance of having the SED shape properly modelled, subsequent studies accounting for the proper SED fitting of other sources will strengthen the framework presented in this paper.

\section{Discussions}
\label{sec:discussions}

As briefly mentioned in the earlier sections, the assumption of the covering factor is perhaps the only weakness in the current model. This value is important to estimate the EW for the respective emission lines. Broad-band SED modelling that includes the torus properties can allow constraining this parameter, either through the study of individual sources \citep[e.g.,][]{Moretal2009} or from large surveys utilizing the optical and infrared fluxes as a proxy for the covering factors \citep[e.g.,][]{Roseboom_etal2013}. Another effective way would be to estimate this parameter using dynamical modelling of the BLR \citep[e.g.,][]{Pancoast_etal2014,Li_etal2016}.

Next, is the issue of constraining the SED through robust modelling and high-quality contemporaneous spectroscopic measurements across the optical, ultraviolet and X-ray energies as has been done for \textmyfont{NGC 5548} and a few other sources \citep[see e.g.,][]{kubota2019,ferlandetal2020}. We need to test the viewing angle-dependent SEDs \citep[e.g.,][]{Wang2014} and compare the modelled predictions to infer physical conditions of the BLR more appropriately.

The location of the dust sublimation radius affects the results obtained in this work. Our assumptions are also supported by \citet{suganuma_2006} which found lags of hot dust emission in AGNs to be $\sim$3.5 times the lag of \hb{} (see their Figure 32a). They thus confirm that region with species like O$^{\circ}$, Mg$^{+}$, Ca$^{+}$ and Fe$^{+}$ lies just inside the hot dust, and in the very outermost part of the BLR. This inference is also validated from our results obtained in this work and earlier in \citet{panda_cafe1,panda2021}.

Another important aspect of the work is the reliability of the spectral fitting techniques and the inference of the \rfe{}. Tests with refined templates \citep[e.g.,][]{2021arXiv211115118P, 2021Univ....7..484M} needs to made to constrain the \rfe{} better for the available sources with high S/N spectroscopy. These then need to be compared with better photoionization models that include \feii{} databases including higher number of transitions \citep[e.g.,][]{sarkar2020}. Our results have shown that whatever is causing \hb{} to vary is also similarly causing \feii{} to vary as well. One can also see in Figure 5 of \citealt{gaskell_2021_FeII} that both \hb{} and \feii{} track the broad features of the continuum variability suggesting similar origins with subtle differences. Better \feii{}-based reverberation mapping estimates are needed to constrain the location of the \feii{}-emitting region in the BLR. This is located further away from the central engine compared to the \hb{} - by a factor $\sim$2 \citep{gaskell_2021_FeII, gaskell_2007_GKN} that is confirmed by the high-cadence reverberation mapping results from \citet{barth_2013, hu15}. Another interesting aspect is the ``breathing'' mode that has been seen especially in the Balmer lines \citep{korista_goad_2004,Barth_etal_2015, runco_etal_2016, gaskell_2021_anomalous}. The variability pattern in the Balmer lines, also studied in the MgII \citep[see e.g.,][]{Guo_etal_2020}, indicate that the location of the onset of the BLR (\rblr{}) can change due to the increase/decrease in the intrinsic luminosity of the source. In order to study this effect and incorporate into our formalism, we need to systematically prepare broad-band SEDs that are representative of such varied epochs in a source. This requires a wide coverage in wavelength, spanning from the optical to X-rays, in addition to AGN continuum light curves. A combination of the two can allow us to test the implication of the breathing mode in terms of the systematic shift in the location of the source in terms of the \un{}.

On the other hand, the \RL{} relation needs to be tested with the inclusion of more reverberation mapped sources spanning the extent of the continuum luminosity. Better proxies of the accretion rate (or \LLEdd{}) are now available, e.g., the near-infrared \caii{} triplet emitting at $\lambda$8498\AA, $\lambda$8542\AA\, and $\lambda$8662\AA \citep{panda_cafe1,2021POBeo.100..287M,martinez-aldama_etal2021}. The prospects for this channel will only get better with the upcoming James Webb Space Telescope and other ground-based observatories, e.g., the Maunakea Spectroscopic Explorer \citep{2019BAAS...51g.126M} and, the European Extremely Large Telescope \citep{2015arXiv150104726E}.

Finally, GRAVITY is just starting to resolve the outer BLR for nearby sources, e.g., 3C 273 \citep{sturm2018} and IRAS 09149-6206 \citep{2020arXiv200908463G} using fantastic interferometric capabilities. And with the upcoming upgrade leading to the GRAVITY+, this will only get better providing us with a spectacular angular resolution that will enable us to pinpoint the location of the BLR in nearby AGNs. Yet, currently, the combination of reverberation mapping and photoionization-based results remains the only credible way to infer the location and physical conditions of these media.

\section{Conclusions}
\label{sec:conclusions}

Through this study, we have:

\begin{itemize}

    \item Tested the variation in the low-ionization emitting regions of the BLR, by accounting for the changes in the shape of the ionizing continuum (the SED) and the location of the BLR from the central ionizing source (or \rblr{}) from the reverberation mapping, in the Eigenvector-1 context. We compare the SEDs for a prototypical Population A and Population B source, \textmyfont{I Zw 1} and \textmyfont{NGC 5548}, respectively in our photoionization modelling using {\sc CLOUDY}.
    
    \item Brought together our knowledge of the BLR \RL{} relations \citep{bentz13,du2019} and the photoionization theory into a unified picture. We highlight the importance of the estimation of the bolometric luminosity that is either (a) scaled-up using the monochromatic luminosity at, e.g., 5100 \AA~, with a fixed factor derived using composite SEDs for Type-1 quasars by combining mid-infrared and optical colours \citep{richards2006}; or (b) uses a luminosity-dependent factor derived using theoretical calculations of optically thick, geometrically thin accretion disks, and observed X-ray properties of AGNs \citep{netzer2019}. We incorporate the two widely used \RL{} relations - the classical \citet{bentz13} relation and the \rfe{}-dependent \RL{} from \citet{du2019} in this approach and compare their behaviour with the photoionization models. Additionally, we test the effect of the inclusion of the $\mathcal{D_{H\beta}}$ in our prescription.
    
    \item Tested the dependence of the location of the optical \feii{} and \hb{} emitting region within the dustless BLR for various cloud parameters, namely, the metal content and turbulence within the BLR cloud. We find that for the case of \textmyfont{NGC 5548}, the solar composition is optimal in recovering the flux ratios. While, for the \textmyfont{I Zw 1} case, the successful models require a BLR composition at least 3\zsun{} with an added effect from turbulence within the cloud. This leads to the enhanced \feii{} emission that then matches the observed estimates.
    
    \item Estimated the EWs for the \hb{} and \feii{} from our photoionization models accounting for covering factors that are verified from previous studies \citep{korista_goad_2000,baldwin2004,panda2021}. We identify pair(s) of solutions for the ionization parameter (\u{}) and local BLR density (\n{}) that agree with the observed line EWs for the low-ionization emitting regions of the dustless BLR. This result highlights the shift in the overall \u{}\n{} recovered from our analysis towards lower values (by up to 2 dex) compared to the \u{}\n{} values estimates from the photoionization theory. This confirms our hypothesis that the BLR ``sees'' a different, filtered SED with only a very small fraction ($\sim$1-10\%) that leads to the line emission in the dustless, low-ionization emitting region of the BLR.
    
\end{itemize}


\section*{Conflict of Interest Statement}

The author declares that the research was conducted in the absence of any commercial or financial relationships that could be construed as a potential conflict of interest.

\section*{Author Contributions}

The idea, analysis and writing of the manuscript has been carried by SP.

\section*{Funding}
The project was partially supported by the Polish Funding Agency National Science Centre, project 2017/26/\-A/ST9/\-00756 (MAESTRO  9) and by Conselho Nacional de Desenvolvimento Científico e Tecnológico (CNPq) Fellowship (164753/2020-6).

\section*{Acknowledgments}
SP would like to thank Prof. Bo\.zena Czerny, Prof. Paola Marziani and Dr Mary Loli Mart\'inez-Aldama for fruitful discussions, and to Ms Denimara Dias dos Santos, Dr Murilo Marinello and Prof. Alberto Rodr\'iguez-Ardila for assisting with the continuum extraction of the \textmyfont{I Zw 1} continuum. The numerical computations have been performed and analyzed using the supercomputing facility at the Nicolaus Copernicus Astronomical Center.


\section*{Data Availability Statement}
The setup and the results from the photoionization simulations using {\sc CLOUDY} can be made available upon request to the author.


\bibliographystyle{frontiersinSCNS_ENG_HUMS} 
\bibliography{references}

\newpage
\appendix

\section{Including the $\mathcal{D_{\rm H\beta}}$ parameter}
\label{app:D-parameter-paramterization}

Another interesting bi-variate relation is found between the \LLEdd{} which is a function of (a) the shape of the broad-line profile ($\mathcal{D}_{H\beta} = FWHM/\sigma_{H\beta}$), and (b) the parameter \rfe{}. The $\mathcal{D}_{\rm H\beta}$ is better known as the line-width ratio \citep[see e.g.,][]{kollatschny_zetzl_2013}, i.e. the ratio between the full-width at half maximum (FWHM) of the emission line (here, \hb{}) and the corresponding dispersion in the \hb{} line ($\sigma_{\rm H\beta}$). In \citet{dupu2016L}, the authors provided a novel relation that is stated as the fundamental relation for the BLR and has the form,

\begin{equation}
    \log \lambda_{\rm{Edd}} = \alpha ' + \beta\mathcal{D}_{H\beta} + \gamma R_\mathrm{\textmyfont{Fe II}}
    \label{eq:blr-fund}
\end{equation} 
The values taken by the coefficients are: $\alpha '$ = 0.31 $\pm$ 0.30, $\beta$ = -0.82 $\pm$ 0.11, and $\gamma$ = 0.80 $\pm$ 0.20.

The monochromatic luminosity at 5100\AA\ (\l5100{}) can be re-written using the black hole mass (\mbh{}) and the Eddington ratio, i.e., $\lambda_{\rm{Edd}}$ = \LLEdd{}. Again, we have two versions with the inclusion of the bolometric correction factor. Assuming the fixed \kbol{} = 9.26 \citep{richards2006}, we have

\begin{equation}
\log (\rm{L_{5100}}) = \log \left(\frac{\rm{L_{bol}}}{9.26}\right) = \log \left(\frac{\rm{\lambda_{Edd}L_{Edd}}}{9.26}\right) = \log (\rm{\lambda_{Edd}}) + \log (\rm{M_{BH}}) + 37.13375956
\label{eq:kbol_const_w_edd}
\end{equation}

While on the other hand assuming the variable \kbol{} \citep{netzer2019}, we have
\begin{equation}
\log (\rm{L_{5100}}) = \log \left(\frac{\rm{L_{bol}}}{k_{bol}}\right) = 1.25\left[\log (\rm{\lambda_{Edd}}) + \log (\rm{M_{BH}})\right] + 35.12288819
\label{eq:kbol_var_w_edd}
\end{equation}

Incorporating this BLR fundamental plane relation in our previously obtained analytical relations with $\log \rm{Un_{H}}$ we get the following relations:


Applying Equation \ref{eq:blr-fund} in Equation \ref{eq5} with Equation \ref{eq:kbol_const_w_edd}, we have
\begin{equation}
    \boxed{\log (Un_{H}) = 9.694 - 0.084\log\left(\frac{\rm{M_{BH}}}{10^8\;\rm{M_{\odot}}}\right) + 0.069\mathcal{D_{\rm H\beta}} - 0.067\rm{R_{Fe II}}}
\end{equation}


Applying Equation \ref{eq:blr-fund} in Equation \ref{eq6} with Equation \ref{eq:kbol_var_w_edd}, we have
\begin{equation}
    \boxed{\log (Un_{H}) = 9.621 - 0.355\log\left(\frac{\rm{M_{BH}}}{10^8\;\rm{M_{\odot}}}\right) + 0.291\mathcal{D_{\rm H\beta}} - 0.284\rm{R_{Fe II}}}
\end{equation}


Applying Equation \ref{eq:blr-fund} in Equation \ref{eq8} with Equation \ref{eq:kbol_const_w_edd}, we have
\begin{equation}
    \boxed{\log (Un_{H}) = 9.769 + 0.1\log\left(\frac{\rm{M_{BH}}}{10^8\;\rm{M_{\odot}}}\right) - 0.082\mathcal{D_{\rm H\beta}} + 0.78\rm{R_{Fe II}}}
\end{equation}


Applying Equation \ref{eq:blr-fund} in Equation \ref{eq9} with Equation \ref{eq:kbol_var_w_edd}, we have
\begin{equation}
    \boxed{\log (Un_{H}) = 9.709 - 0.125\log\left(\frac{\rm{M_{BH}}}{10^8\;\rm{M_{\odot}}}\right) + 0.103\mathcal{D_{\rm H\beta}} + 0.6\rm{R_{Fe II}}}
\end{equation}

The estimates from the above analytical relations are summarised in Table \ref{tab:table3}. Similar to Table \ref{tab:table1}, we report the estimates for all the cases accounting for the appropriate $\chi$ values for the two sources in the last two columns in Table \ref{tab:table3}.

\begin{table}[!htb]
\centering
\caption{Estimates for log(Un$_{\rm H}$) for the various relations considered in this paper including the $\mathcal{D_{\rm H\beta}}$ parameter}
\label{tab:table3}
\resizebox{\textwidth}{!}{%
\setlength{\tabcolsep}{0.5em} 
\renewcommand{\arraystretch}{2}
\begin{tabular}{c|c|c|c|c|c|c}
\hline
\textbf{Radius-Luminosity relation}  & \textbf{Bolometric Correction} & \textbf{log(Un$_{\rm H}$)}                           & \textbf{NGC 5548}$^{a}$               & \textbf{I Zw 1}$^{b}$   & \textbf{NGC 5548}$^{c}$               & \textbf{I Zw 1}$^{d}$                  \\ \hline
\multirow{2}{*}{\citet{bentz13}} & \citet{richards2006}        & 9.694 - 0.084log$\left(\frac{\rm{M_{BH}}}{10^8\;\rm{M_{\odot}}}\right)$ + 0.069$\mathcal{D_{\rm H\beta}}$ - 0.067\rfe{}                   & 9.895                    & 9.647  & 10.110                    & 9.027                     \\ \cline{2-7} 
                                     & \citet{netzer2019}                  & 9.621 - 0.355log$\left(\frac{\rm{M_{BH}}}{10^8\;\rm{M_{\odot}}}\right)$ + 0.291$\mathcal{D_{\rm H\beta}}$ - 0.284\rfe{}                   & 10.467                     & 9.420 & 10.682                     & 8.800                     \\ \hline
\multirow{2}{*}{\citet{du2019}}   & \citet{richards2006}        & 9.769 + 0.1log$\left(\frac{\rm{M_{BH}}}{10^8\;\rm{M_{\odot}}}\right)$ - 0.082$\mathcal{D_{\rm H\beta}}$ + 0.78\rfe{}  & 9.600  & 10.959 & 9.815  & 10.339 \\ \cline{2-7} 
                                     & \citet{netzer2019}                  & 9.709 - 0.125log$\left(\frac{\rm{M_{BH}}}{10^8\;\rm{M_{\odot}}}\right)$ + 0.103$\mathcal{D_{\rm H\beta}}$ + 0.6\rfe{} & 10.078 & 10.772    & 10.293 & 10.152 \\ \hline
\end{tabular}%
}
{\footnotesize
$^{@}$ denotes for the case with $\chi$=0.5 which is used to estimate values for {\sc NGC 5548} and {\sc I Zw 1} in columns 4 and 5, respectively. Black hole mass (\mbh{}) for: $^{a}$ \textmyfont{NGC 5548} = (5.2$\pm$1.3)$\times$10$^7$ \msun{} \citep{fausnaughetal2016}; $^{b}$ \textmyfont{I Zw 1} = (8.61$\pm$1.50)$\times$10$^7$ \msun{} \citep[][]{martinez-aldama_etal2021}. The \mbh{} is estimated using the spectral properties obtained from \citet{persson1988}. These are consistent with their respective SEDs considered in this paper. The corresponding $\mathcal{D_{\rm H\beta}}$ and \rfe{} for (a) \textmyfont{NGC 5548} are 2.66$\pm$0.13, and 0.1$\pm$0.02 \citep{du2019}, respectively; and for (b) \textmyfont{I Zw 1} are 0.809$\pm$0.13, and 1.619$\pm$0.060 (Marziani et al. 2021, submitted), respectively. $^{c}$ uses the $\chi$=0.82 as reported by {\sc CLOUDY} for {\sc NGC 5548}, $^{d}$ uses the $\chi$=0.12 as reported by {\sc CLOUDY} for {\sc I Zw 1}, keeping other parameters identical as before.}
\end{table}

\begin{figure}[!htb]
    \centering
    \includegraphics[width=\textwidth]{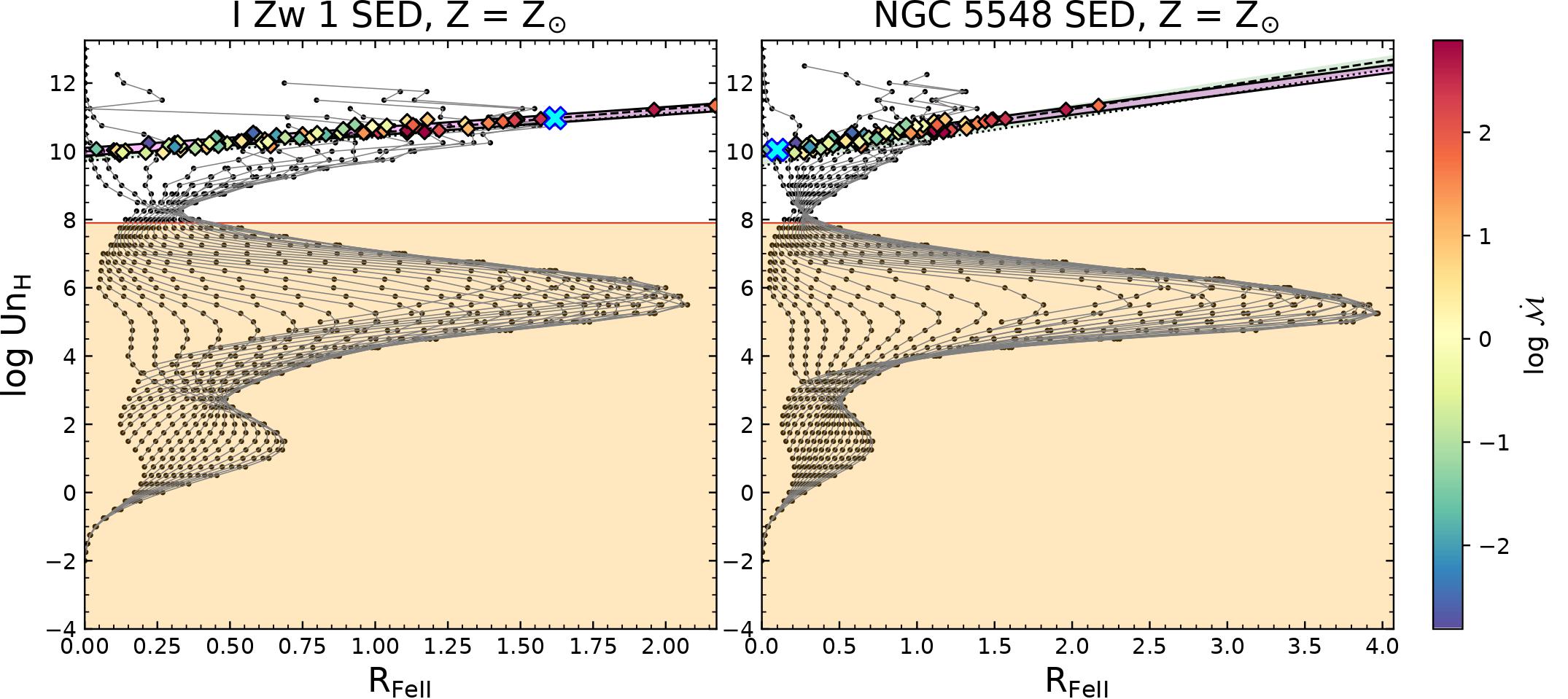}
    \caption{Similar to Figure \ref{fig:grid_IZw1_NGC5548_wo_D} but with the inclusion of the $\mathcal{D_{\rm H\beta}}$ in the analytical prescription. The panels include the full range of the $\mathcal{D_{\rm H\beta}}$ as per min-max range from the histogram shown in Figure \ref{fig:D-RFe-histogram-Du2019}. The \u{}\n{} estimates for the observed sample is based on the luminosity-dependent \kbol{} and \rfe{}-based \RL{} relation as shown in Table \ref{tab:table3}. The blue cross in each panel shows the location of the respective sources from the estimates tabulated in the last row of Table \ref{tab:table3}.}
    \label{fig:grid_IZw1_NGC5548_w_D}
\end{figure}

\newpage
\begin{longtable}[c]{lccccc}
\caption{Spectral and derived properties of reverberation-mapped AGNs}
\label{tab:rm-data}\\
\hline
\multirow{2}{*}{Source} & log L$_{5100}$ & log M$_{\rm BH}$ & \multirow{2}{*}{log $\mathcal{\dot{M}}$} & \multirow{2}{*}{R$_{\rm FeII}$} & \multirow{2}{*}{$\mathcal{D}_{H\beta}$} \\
 & {[}erg s$^{-1}${]} & {[}M$_{\odot}${]} &  &  &  \\ \hline
\endfirsthead
\endhead
\hline
\endfoot
\endlastfoot
Mrk 335 & 43.76$\pm$0.07 & 6.93$^{+0.1}_{-0.11}$ & 1.27$^{+0.18}_{-0.17}$ & 0.62$\pm$0.124 & 1.27$\pm$0.05 \\
PG 0026+129 & 44.97$\pm$0.02 & 8.15$^{+0.09}_{-0.13}$ & 0.65$^{+0.28}_{-0.2}$ & 0.33$\pm$0.066 & 1.46$\pm$0.09 \\
PG 0052+251 & 44.81$\pm$0.03 & 8.64$^{+0.11}_{-0.14}$ & -0.59$^{+0.31}_{-0.25}$ & 0.12$\pm$0.024 & 2.31$\pm$0.05 \\
Fairall9 & 43.98$\pm$0.04 & 8.09$^{+0.07}_{-0.12}$ & -0.71$^{+0.31}_{-0.21}$ & 0.49$\pm$0.098 & 2.56$\pm$0.03 \\
Mrk 590 & 43.5$\pm$0.21 & 7.55$^{+0.07}_{-0.08}$ & -0.41$^{+0.36}_{-0.36}$ & 0.45$\pm$0.09 & 1.39$\pm$0.07 \\
Mrk 1044 & 43.1$\pm$0.1 & 6.45$^{+0.12}_{-0.13}$ & 1.22$^{+0.4}_{-0.41}$ & 0.99$\pm$0.198 & 1.54$\pm$0.03 \\
3C 120 & 44$\pm$0.1 & 7.79$^{+0.15}_{-0.15}$ & -0.07$^{+0.3}_{-0.3}$ & 0.39$\pm$0.078 & 1.86$\pm$0.05 \\
IRAS 04416+1215 & 44.47$\pm$0.03 & 6.78$^{+0.31}_{-0.06}$ & 2.63$^{+0.16}_{-0.67}$ & 1.96$\pm$0.392 & 1.44$\pm$0.06 \\
Ark 120 & 43.87$\pm$0.25 & 8.47$^{+0.07}_{-0.08}$ & -1.7$^{+0.41}_{-0.41}$ & 0.83$\pm$0.166 & 1.65$\pm$0.01 \\
MCG +08-11-011 & 43.33$\pm$0.11 & 7.72$^{+0.04}_{-0.05}$ & -0.96$^{+0.25}_{-0.28}$ & 0.29$\pm$0.058 & 1.55$\pm$0.11 \\
Mrk 374 & 43.77$\pm$0.04 & 7.86$^{+0.15}_{-0.12}$ & -0.56$^{+0.3}_{-0.36}$ & 0.88$\pm$0.176 & 1.38$\pm$0.1 \\
Mrk 79 & 43.68$\pm$0.07 & 7.84$^{+0.12}_{-0.16}$ & -0.68$^{+0.25}_{-0.21}$ & 0.33$\pm$0.066 & 2.1$\pm$0.06 \\
SDSS J074352 & 45.37$\pm$0.02 & 7.93$^{+0.05}_{-0.04}$ & 1.69$^{+0.12}_{-0.13}$ & 1.11$\pm$0.222 & 1.6$\pm$0.02 \\
SDSS J075051 & 45.33$\pm$0.01 & 7.67$^{+0.11}_{-0.07}$ & 2.14$^{+0.16}_{-0.24}$ & 1.22$\pm$0.244 & 1.54$\pm$0.01 \\
SDSS J075101 & 44.18$\pm$0.09 & 7.18$^{+0.08}_{-0.09}$ & 1.39$^{+0.21}_{-0.21}$ & 0.97$\pm$0.194 & 1.52$\pm$0.08 \\
Mrk 382 & 43.12$\pm$0.08 & 6.5$^{+0.19}_{-0.29}$ & 1.18$^{+0.69}_{-0.53}$ & 0.75$\pm$0.15 & 1.74$\pm$0.36 \\
SDSS J075949 & 44.2$\pm$0.03 & 7.44$^{+0.25}_{-0.25}$ & 0.89$^{+0.41}_{-0.41}$ & 1.02$\pm$0.204 & 1.49$\pm$0.04 \\
SDSS J080101 & 44.27$\pm$0.03 & 6.78$^{+0.34}_{-0.17}$ & 2.33$^{+0.39}_{-0.72}$ & 1.01$\pm$0.202 & 1.61$\pm$0.08 \\
SDSS J080131 & 43.97$\pm$0.04 & 6.51$^{+0.22}_{-0.17}$ & 2.43$^{+0.29}_{-0.36}$ & 1.49$\pm$0.298 & 1.36$\pm$0.03 \\
PG 0804+761 & 44.91$\pm$0.02 & 8.43$^{+0.05}_{-0.06}$ & 0.0$^{+0.15}_{-0.13}$ & 0.61$\pm$0.122 & 2.13$\pm$0.04 \\
SDSS J081441 & 43.96$\pm$0.06 & 7.18$^{+0.2}_{-0.2}$ & 1.09$^{+0.35}_{-0.35}$ & 0.46$\pm$0.092 & 1.54$\pm$0.08 \\
SDSS J081456 & 43.99$\pm$0.04 & 7.44$^{+0.12}_{-0.49}$ & 0.59$^{+1.03}_{-0.3}$ & 1.31$\pm$0.262 & 1.71$\pm$0.09 \\
NGC 2617 & 42.67$\pm$0.16 & 7.74$^{+0.11}_{-0.17}$ & -1.98$^{+0.55}_{-0.51}$ & 0.31$\pm$0.062 & 2.55$\pm$0.18 \\
SDSS J083553 & 44.44$\pm$0.02 & 6.87$^{+0.16}_{-0.25}$ & 2.41$^{+0.53}_{-0.35}$ & 1.57$\pm$0.314 & 1.73$\pm$0.02 \\
SDSS J084533 & 44.53$\pm$0.02 & 6.76$^{+0.14}_{-0.15}$ & 2.76$^{+0.24}_{-0.24}$ & 1.11$\pm$0.222 & 1.38$\pm$0.03 \\
PG 0844+349 & 44.22$\pm$0.07 & 7.66$^{+0.15}_{-0.23}$ & 0.5$^{+0.57}_{-0.42}$ & 0.78$\pm$0.156 & 1.79$\pm$0.04 \\
SDSS J085946 & 44.41$\pm$0.03 & 7.3$^{+0.19}_{-0.61}$ & 1.51$^{+1.27}_{-0.43}$ & 1.39$\pm$0.278 & 1.59$\pm$0.04 \\
Mrk 110 & 43.66$\pm$0.12 & 7.1$^{+0.13}_{-0.14}$ & 0.77$^{+0.26}_{-0.25}$ & 0.14$\pm$0.028 & 1.69$\pm$0.09 \\
SDSS J093302 & 44.31$\pm$0.13 & 7.08$^{+0.08}_{-0.11}$ & 1.79$^{+0.4}_{-0.4}$ & 1.44$\pm$0.288 & 1.26$\pm$0.02 \\
SDSS J093922 & 44.07$\pm$0.04 & 6.53$^{+0.07}_{-0.33}$ & 2.54$^{+0.71}_{-0.2}$ & 1.48$\pm$0.296 & 1.29$\pm$0.06 \\
PG 0953+414 & 45.19$\pm$0.01 & 8.44$^{+0.06}_{-0.07}$ & 0.39$^{+0.16}_{-0.14}$ & 0.27$\pm$0.054 & 1.85$\pm$0.04 \\
SDSS J100402 & 45.52$\pm$0.01 & 7.44$^{+0.37}_{-0.06}$ & 2.89$^{+0.13}_{-0.75}$ & 1.17$\pm$0.234 & 1.47$\pm$0.01 \\
SDSS J101000 & 44.76$\pm$0.02 & 7.46$^{+0.27}_{-0.14}$ & 1.7$^{+0.31}_{-0.56}$ & 2.17$\pm$0.434 & 1.64$\pm$0 \\
NGC 3227 & 42.24$\pm$0.11 & 7.09$^{+0.09}_{-0.12}$ & -1.34$^{+0.38}_{-0.36}$ & 0.46$\pm$0.092 & 2.44$\pm$0.17 \\
SDSS J102339 & 44.09$\pm$0.03 & 7.16$^{+0.25}_{-0.08}$ & 1.29$^{+0.2}_{-0.56}$ & 1.03$\pm$0.206 & 1.48$\pm$0.04 \\
Mrk 142 & 43.59$\pm$0.04 & 6.47$^{+0.38}_{-0.38}$ & 1.93$^{+0.59}_{-0.59}$ & 1.14$\pm$0.228 & 1.36$\pm$0.26 \\
NGC 3516 & 42.79$\pm$0.2 & 7.82$^{+0.05}_{-0.08}$ & -1.97$^{+0.41}_{-0.52}$ & 0.66$\pm$0.132 & 2.45$\pm$0.17 \\
SBS 1116+583A & 42.14$\pm$0.23 & 6.78$^{+0.11}_{-0.12}$ & -0.87$^{+0.51}_{-0.71}$ & 0.59$\pm$0.118 & 2.36$\pm$0.13 \\
Arp 151 & 42.55$\pm$0.1 & 6.87$^{+0.05}_{-0.08}$ & -0.44$^{+0.3}_{-0.28}$ & 0.32$\pm$0.064 & 1.54$\pm$0.04 \\
NGC 3783 & 42.56$\pm$0.18 & 7.45$^{+0.12}_{-0.11}$ & -1.58$^{+0.45}_{-0.59}$ & 0.04$\pm$0.008 & 2.23$\pm$0.05 \\
MCG +06-26-012 & 42.67$\pm$0.11 & 6.92$^{+0.14}_{-0.12}$ & -0.34$^{+0.37}_{-0.45}$ & 1.04$\pm$0.208 & 1.7$\pm$0.11 \\
UGC 06728 & 41.86$\pm$0.08 & 5.87$^{+0.19}_{-0.4}$ & 0.55$^{+0.92}_{-0.51}$ & 1.11$\pm$0.222 & 0.89$\pm$0.12 \\
Mrk 1310 & 42.29$\pm$0.14 & 6.62$^{+0.07}_{-0.08}$ & -0.31$^{+0.35}_{-0.39}$ & 0.46$\pm$0.092 & 1.99$\pm$0.07 \\
NGC 4051 & 41.9$\pm$0.15 & 5.72$^{+0.34}_{-0.44}$ & 0.9$^{+0.79}_{-0.74}$ & 1.18$\pm$0.236 & 1.97$\pm$0.15 \\
NGC 4151 & 42.09$\pm$0.21 & 7.72$^{+0.07}_{-0.06}$ & -2.81$^{+0.37}_{-0.57}$ & 0.22$\pm$0.044 & 2.76$\pm$0.07 \\
PG 1211+143 & 44.73$\pm$0.08 & 7.87$^{+0.11}_{-0.26}$ & 0.84$^{+0.63}_{-0.35}$ & 0.42$\pm$0.084 & 1.35$\pm$0.04 \\
Mrk 202 & 42.26$\pm$0.14 & 6.11$^{+0.2}_{-0.2}$ & 0.66$^{+0.59}_{-0.65}$ & 0.57$\pm$0.114 & 1.7$\pm$0.08 \\
NGC 4253 & 42.57$\pm$0.12 & 6.49$^{+0.1}_{-0.1}$ & 0.36$^{+0.36}_{-0.42}$ & 0.99$\pm$0.198 & 1.48$\pm$0.06 \\
PG 1226+023 & 45.92$\pm$0.05 & 8.5$^{+0.03}_{-0.04}$ & 1.37$^{+0.15}_{-0.14}$ & 0.64$\pm$0.128 & 1.97$\pm$0.03 \\
PG 1229+204 & 43.7$\pm$0.05 & 8.03$^{+0.24}_{-0.23}$ & -1.03$^{+0.52}_{-0.55}$ & 0.53$\pm$0.106 & 2.38$\pm$0.05 \\
NGC 4593 & 42.62$\pm$0.37 & 7.26$^{+0.09}_{-0.09}$ & -1.1$^{+0.6}_{-0.64}$ & 0.89$\pm$0.178 & 2.87$\pm$0.66 \\
IRAS F12397+3333 & 44.23$\pm$0.05 & 6.79$^{+0.27}_{-0.45}$ & 2.26$^{+0.98}_{-0.62}$ & 1.48$\pm$0.296 & 1.57$\pm$0.52 \\
NGC 4748 & 42.56$\pm$0.12 & 6.61$^{+0.11}_{-0.23}$ & 0.1$^{+0.61}_{-0.44}$ & 0.99$\pm$0.198 & 1.93$\pm$0.08 \\
PG 1307+085 & 44.85$\pm$0.02 & 8.72$^{+0.13}_{-0.26}$ & -0.68$^{+0.53}_{-0.28}$ & 0.21$\pm$0.042 & 2.58$\pm$0.09 \\
MCG +06-30-015 & 41.64$\pm$0.11 & 6.63$^{+0.12}_{-0.15}$ & -1.29$^{+0.37}_{-0.38}$ & 0.93$\pm$0.186 & 2.01$\pm$0.08 \\
NGC 5273 & 41.54$\pm$0.16 & 7.14$^{+0.19}_{-0.56}$ & -2.5$^{+1.33}_{-0.67}$ & 0.58$\pm$0.116 & 3.12$\pm$0.13 \\
Mrk 279 & 43.71$\pm$0.07 & 7.97$^{+0.09}_{-0.12}$ & -0.89$^{+0.33}_{-0.3}$ & 0.55$\pm$0.11 & 2.94$\pm$0.03 \\
PG 1411+442 & 44.56$\pm$0.02 & 8.28$^{+0.17}_{-0.3}$ & -0.23$^{+0.63}_{-0.38}$ & 0.63$\pm$0.126 & 1.58$\pm$0.04 \\
NGC 5548 & 43.3$\pm$0.19 & 8.08$^{+0.16}_{-0.16}$ & -1.76$^{+0.31}_{-0.32}$ & 0.1$\pm$0.02 & 2.66$\pm$0.33 \\
PG 1426+015 & 44.63$\pm$0.02 & 8.97$^{+0.12}_{-0.22}$ & -1.51$^{+0.47}_{-0.28}$ & 0.46$\pm$0.092 & 2.45$\pm$0.09 \\
Mrk 817 & 43.74$\pm$0.09 & 7.99$^{+0.14}_{-0.14}$ & -0.87$^{+0.22}_{-0.22}$ & 0.69$\pm$0.138 & 2.59$\pm$0.29 \\
Mrk 1511 & 43.16$\pm$0.06 & 7.29$^{+0.07}_{-0.07}$ & -0.34$^{+0.24}_{-0.24}$ & 0.8$\pm$0.16 & 2.2$\pm$0.13 \\
Mrk 290 & 43.17$\pm$0.06 & 7.55$^{+0.07}_{-0.07}$ & -0.85$^{+0.23}_{-0.23}$ & 0.29$\pm$0.058 & 2.57$\pm$0.18 \\
Mrk 486 & 43.69$\pm$0.05 & 7.24$^{+0.12}_{-0.06}$ & 0.55$^{+0.2}_{-0.32}$ & 0.54$\pm$0.108 & 1.5$\pm$0.06 \\
Mrk 493 & 43.11$\pm$0.08 & 6.14$^{+0.04}_{-0.11}$ & 1.88$^{+0.33}_{-0.21}$ & 1.13$\pm$0.226 & 1.52$\pm$0.03 \\
PG 1613+658 & 44.77$\pm$0.02 & 8.81$^{+0.14}_{-0.21}$ & -0.97$^{+0.45}_{-0.31}$ & 0.38$\pm$0.076 & 2.94$\pm$0.05 \\
PG 1617+175 & 44.39$\pm$0.02 & 8.79$^{+0.15}_{-0.28}$ & -1.5$^{+0.58}_{-0.33}$ & 0.74$\pm$0.148 & 2.87$\pm$0.12 \\
PG 1700+518 & 45.59$\pm$0.01 & 8.4$^{+0.08}_{-0.08}$ & 1.08$^{+0.17}_{-0.17}$ & 1.32$\pm$0.264 & 1.09$\pm$0.08 \\
3C 382 & 43.84$\pm$0.1 & 8.67$^{+0.09}_{-0.06}$ & -2.09$^{+0.26}_{-0.35}$ & 0.31$\pm$0.062 & 1.87$\pm$0.13 \\
3C 390.3 & 44.43$\pm$0.58 & 9.18$^{+0.23}_{-0.23}$ & -2.62$^{+0.95}_{-0.96}$ & 0.12$\pm$0.024 & 2.62$\pm$0.66 \\
KA 1858+4850 & 43.43$\pm$0.05 & 6.94$^{+0.07}_{-0.09}$ & 0.75$^{+0.25}_{-0.21}$ & 0.11$\pm$0.022 & 2.13$\pm$0.13 \\
NGC 6814 & 42.12$\pm$0.28 & 7.16$^{+0.05}_{-0.06}$ & -1.64$^{+0.46}_{-0.8}$ & 0.45$\pm$0.09 & 1.73$\pm$0.03 \\
Mrk 509 & 44.19$\pm$0.05 & 8.15$^{+0.03}_{-0.03}$ & -0.52$^{+0.13}_{-0.14}$ & 0.13$\pm$0.026 & 1.94$\pm$0.01 \\
PG 2130+099 & 44.32$\pm$0.04 & 7.29$^{+0.06}_{-0.09}$ & 1.4$^{+0.24}_{-0.19}$ & 0.96$\pm$0.192 & 1.39$\pm$0.11 \\
NGC 7469 & 43.51$\pm$0.11 & 7.6$^{+0.12}_{-0.06}$ & -0.46$^{+0.26}_{-0.42}$ & 0.43$\pm$0.086 & 1.49$\pm$0.18 \\ \hline
\end{longtable}
\vspace{-0.75cm}
\begin{longtable}{c}

\vspace{-0.15cm}
\footnotesize{Columns\footnote{Compiled using the estimates provided in Tables 1 (electronically available also on \href{https://cdsarc.cds.unistra.fr/viz-bin/cat/J/ApJ/886/42}{CDS/Vizier})  and 2 in \citet{du2019}.} depict: (1) Source name; (2) luminosity at 5100\AA~ in log-scale; (3) black hole mass in log-scale; (3) dimensionless accretion rate ($\mathcal{\dot{M}}$) in log-scale;}\\
\vspace{-0.5cm}
\footnotesize{(4) the ratio of the EW(\feii{}) in the optical within 4434-4684\AA~ to the EW(\hb{}); and (5) ratio of the FWHM(\hb{}) to the $\sigma_{H\beta}$.}
\end{longtable}
\newpage

\section{Supplementary plots from {\sc CLOUDY} simulations}

\begin{figure}[!htb]
    \centering
    \includegraphics[width=\textwidth]{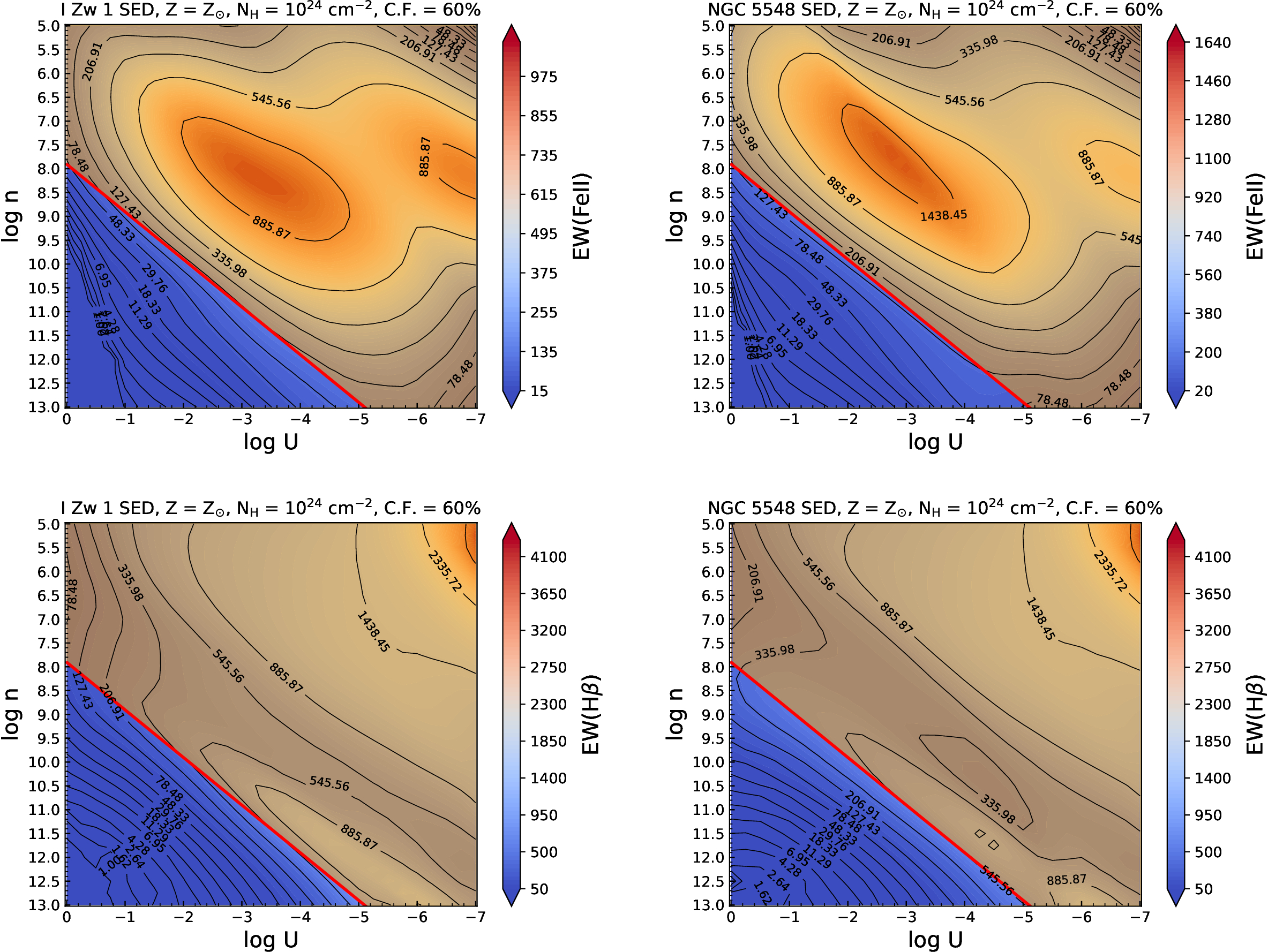}
    \caption{\un{} 2D histograms color-weighted by (top panels) equivalent width (EWs) \feii{}, and (bottom panels) EW(\hb{}). The labels and parameters shown here are identical to Figure \ref{fig:compare_IZw1_NGC5548_2dhist_EW_20pc} except here we use a larger value for the covering fraction, i.e., 60\%.}
    \label{fig:compare_IZw1_NGC5548_2dhist_EW_60pc}
\end{figure}

\end{document}